\documentclass[aps,prb,twocolumn,groupedaddress,amsmath,amssymb,showpacs,floatfix]{revtex4-1}
\usepackage{graphics,color,epsfig}
\graphicspath{{./rev_plots/}}
\usepackage{float}
\usepackage{appendix}
\usepackage{amsmath}
\usepackage{amssymb}
\usepackage{amsfonts}
\usepackage{amsfonts}
\usepackage{epstopdf}
\usepackage{bm}
\usepackage{times,xspace}
\usepackage{color}
\usepackage[utf8]{inputenc}
\usepackage{multirow}
\usepackage{array}
\usepackage{tabularx}

\def\beq{\begin{eqnarray}}\def\eeq{\end{eqnarray}}
\def\be{\begin{equation}}\def\ee{\end{equation}}

\def\T{${\mathcal{T}}$}
\def\P{${\mathcal{P}}$}
\def\C{${\mathcal{C}}$}

\def\CP{${\mathcal{CP}}$}
\def\CT{${\mathcal{CT}}$}
\def\PT{${\mathcal{PT}}$}
\def\CPT{${\mathcal{CPT}}$}
\def\Z{${\mathbb{Z} }$}
\def\ZZ{${\mathbb{Z}_2 }$}

\newcommand{\mcz}[0]{\mathbb{Z}}
\newcommand{\mczz}[0]{\mathbb{Z}_2}
\newcommand{\mcp}[0]{\mathcal{P}}
\newcommand{\mct}[0]{\mathcal{T}}
\newcommand{\mcc}[0]{\mathcal{C}}

\newcommand{\mck}[0]{\mathcal{K}}
\newcommand{\mcct}[0]{\mathcal{CT}}
\newcommand{\mcpt}[0]{\mathcal{PT}}
\newcommand{\mccp}[0]{\mathcal{CP}}
\newcommand{\mccpt}[0]{\mathcal{CPT}}

\newcommand{\cc}[0]{\mathcal{K}}

\newcommand{\g}[0]{\Gamma}
\newcommand{\z}[0]{\mathbb{Z}}
\newcommand{\zz}[0]{\mathbb{Z}_2}

\begin{document}

\title{Rigidity of topological invariants to symmetry breakings}

\author{Arpit Raj, Nepal Banerjee, and Tanmoy Das}
\email{tnmydas@iisc.ac.in}
\affiliation{Department of Physics, Indian Institute of Science, Bangalore-560012, India}

\date{\today}

\begin{abstract}
Symmetry plays an important role in the topological band theory to remedy the eigenstates' gauge obstruction at the cost of a symmetry anomaly and zero-energy boundary modes. One can also make use of the symmetry to enumerate the topological invariants $-$ giving a symmetry classification table. Here we consider various topological phases protected by different symmetries, and examine how the corresponding topological invariants evolve once the protecting symmetry is spontaneously lost. To our surprise, we find that the topological invariants and edge states can sometimes be robust to symmetry breaking quantum orders. This topological robustness persists as long as the mean-field Hamiltonian in a symmetry breaking ordered phase maintains its adiabatic continuity to the non-interacting Hamiltonian. For example, for a time-reversal symmetric topological phase in 2+1D, we show that the $\mathbb{Z}_2$ time-reversal polarization continues to be a good topological invariant even after including distinct time-reversal breaking order parameters. Similar conclusions are drawn for various other symmetry breaking cases. Finally, we discuss that the change in the internal symmetry associated with the spontaneous symmetry breaking has to be accounted for to reinstate the topological invariants into the expected classification table.
\end{abstract}
\pacs{Topological phases, Topological phase transition, Spontaneous symmetry breaking}
\maketitle

\section{Introduction}

It's more than two decades now that the concept of topology, which plays an important role in the gauge theory in QED and QCD, has found a place in the electronic structure of condensed matters.\cite{Shifmanbook, WeinbergBook,Nakahara, WittenRMP,DasRMP} Within the electronic band theory, topology arises intrinsically due to the presence of a Berry gauge connection $-$ a consequence of the obstruction to the smooth (global) gauge fixing in the wavefunction in the Brillouin zone (BZ).\cite{SCZTQFT,KaneRMP,SCZRMP,Bernevigbook} To reinstate the global symmetry, one requires to loose one or more discrete symmetry (s). This symmetry is actually invariant in the classical theory, but is broken at the quantum level $-$ hence the term symmetry/quantum \textit{anomaly}.\cite{Shifmanbook,Nakahara,Fujikawa,ABJAnomaly,tHooft,Nielsen} The symmetry anomaly results from a one-way \textit{spectral flow} between the states that transform under the symmetry $-$ dubbed \textit{symmetry partners}. The anomaly splits the symmetry partners to be separately localized at different boundaries as zero-energy modes, but remain connected via bulk insulating bands.\cite{foot1} The one-way spectral flow between these symmetry partners at the boundaries, via the bulk states, causes a symmetry-dictated polarization such as charge pumping for the chiral anomaly,\cite{ThoulessPump,LaughlinArg} spin/helicity pumping for the TR polarization \cite{KaneMele,SCZQSH,DasRMP,KaneRMP,SCZRMP}, Majorana pumping for the charge conjugation anomaly.\cite{KitaevMajorana,TopSCSato}

Accordingly, discrete symmetries play important roles to the topological protections. In fact, it is soon realized that all the topological phases can be classified by discrete symmetries.\cite{AZ,Schnyder,KitaevPerTab,TenfoldRMP,TenfoldOthers,TCIClass,TIClassReflection,TIClassSpace,TIClassClifford} Discrete symmetries help to uniquely organize the mappings from eigenspaces to corresponding homotopy groups, and enumerate the topological invariants by the winding numbers of the homotopy group belonging to either the \Z~ or \ZZ~ or 0 (trivial) phases. Relevant discrete symmetries include charge conjugation (CC, denoted by operator \C), parity (operator \P), time-reversal (TR, with operator \T), and others. The combination of CC, TR, and their product (\CT), which gives the chiral or sublattice symmetries, forms a closed group algebra. The corresponding topological phases can be classified by these symmetries, forming the so-called \textit{ten-fold} classification scheme.\cite{AZ,Schnyder,KitaevPerTab,TenfoldRMP,TenfoldOthers} Similarly, one can generalize the classifications to other discrete symmetries, say, point-group, space group, and non-symmorphic symmetries etc., forming various distinct classification schemes. \cite{TCIClass,TIClassReflection,TIClassSpace,TIClassClifford}

Here we raise these questions: What happens when the protecting/anomalous symmetry is spontaneously (or explicitly) broken? How do the bulk topological invariants and the gapless edge modes respond to the loss of these symmetries?  Does the topological invariant shift its position in the \textit{ten-fold} classification table?    

The question becomes more interesting when the symmetry breaking order parameter, at least at the mean-field level, does not close the original topological band (rather quasiparticle) gap, and also does not modify the value of the corresponding topological invariant. Our findings here reveal that although a change in symmetry dictates a transition from one homotopy group to another with different winding numbers, the topological invariant continues to belong to its original non-interacting homotopy group. Furthermore, we also find that the corresponding symmetry-dictated polarization, such as charge/spin/TR polarization as appropriate $-$ an indicator of the symmetry anomaly $-$ remains a good indicator of the topological phase even after the loss of the corresponding symmetry. For example, the so-called \ZZ~ invariant or spin Chern number, which is a measure of the TR polarization in \T$^2=-1$ invariant systems,\cite{KaneMele, FuKaneMele, FuKane,SCZQSH} maintains is non-trivial value even after the TR symmetry is spontaneously broken. Finally, we discuss these results in terms of the change in the internal symmetry and the representation of the eigenstates associated with the spontaneous symmetry breaking, which may be responsible for the change in the winding number of the homotopy group. 

We demonstrate the above phenomena based on a model Hamiltonian in 2+1 D. The non-interacting Hamiltonian is designed in a bilayer of 2D electron gas. The Rashba-type spin-orbit coupling (SOC) is reversed in sign between the layers. The system attains a topological phase as the inter-layer coupling is tuned above a critical value. This coupling becomes the topological band gap (Dirac mass). The Hamiltonian features a \ZZ~ topological phase protected by the $\mct^2=-1$ symmetry, belonging to the DIII topological phases according to the \textit{ten-fold~way} table. The results are affirmed by both the explicit calculations of \ZZ~ TR polarization index $\nu$ and the spin Chern number $C_s$.

Next we consider various symmetry breaking order parameters within an extended Hubbard model. Different order parameters break different symmetries, such as \C, \T, \P~ and their various dual and trinal combinations.

(i) First we consider a sublattice density (SD) order that breaks \C, \P~symmetries, but preserves \CP~ and \T, and hence has the \CPT~ symmetry.\cite{footCPT} Interestingly, the system continues to possess the same \ZZ~topological invariant $\nu\in \mczz$ (but not the spin Chern number), and the gapless Dirac cone at the boundary. This phase is adiabatically connected to the non-interacting quantum spin Hall (QSH) phase, however, the topological class changes its place from the DIII to the AII cartan group $-$ both maps to the same \ZZ~homotopy group. 

(ii) Next, we consider a ferromagnetic (FM) order. Such a state respects \C, \P, and \CP~ symmetries but lacks the \CPT~ invariance. The symmetry classification now switches to the D-class, which, according to the \textit{ten-fold way} table, maps to the \Z~ homotopy group. However, the mean-field FM Hamiltonian continues to feature the same spin Chern number $C_s$ and TR invariant $\nu\in \mczz$ as in the parent QSH phase, as long as the FM exchange energy is lower than the original Dirac mass. Finally, above a critical value of the FM exchange energy, a topological phase transition occurs to the expected \Z~group. 

(iii) We also consider an inter-sublattice magnetic (SM) order which breaks \T, \C~ symmetries, but respects the sublattice (\CT), and $(\mccpt)^2=-1$ symmetries.  Most interestingly, in this case the TR breaking perturbation \textit{never} destroys the \ZZ~ invariance and the spin Chern number. The Dirac cones at the boundaries remain gapless but the gapless points move away from the TR invariant ${\bf k}=0$ point.

(iv) We further consider an antiferromagnetic (AF) order, which breaks \P~, but recovers the $(\mcpt)^2=-1$ invariance. This gives a \ZZ~topological invariant $\nu=1$ in the bulk and the system remains adiabatically connected to the non-interacting bands. But the edge states are now gapped. Here we also find that the AF gap, and the non-interacting interlayer hopping amplitude conspire to yield a complex tunneling term. This complex tunneling term gives a $O(2)$ field and a vortex-like structure. The winding number associated with the vortex corresponds to the same \ZZ~topological invariant. 

(v) We introduce a novel sublattice vortex (SV) phase which breaks different set of symmetries, yet, gives all the same topological invariants as the AF phase does. 

Finally, we motivate a discussion of our results in Sec.~\ref{Sec:Discussion} mainly in two directions. (i) The  winding number of the  $d^{\rm th}$-homotopy $\pi_d(SU(N))$ depends on the value of $N$ for a given $d$. Since the spontaneous symmetry breaking occurs in the wavefunction, it may change the internal symmetry $SU(N)$ and the irreducible representation of the eigenstates depending on the system at hand. (ii) Secondly, we  discuss how symmetries are sometimes treated differently in the classical and quantum theories. For example, the symmetry anomaly is a contradiction in which the classical theory is invariant under the symmetry, but not its quantum version. We cast our results as a collection of counter-examples to the symmetry anomaly. We find that if the symmetry in the classical theory is spontaneously broken, the quantum theory, in which the symmetry is already anomalous, may not be immediately sensitive to it as long as the bulk topological gap is not closed. 

\section{A prelude}\label{Sec:Prelude}
To direct the discussions on the topological phases and discrete symmetries into our research topics, we start with a brief account on how they are connected. We discuss how the topological classifications are performed based on the obstruction to the gauge fixing for complex eigenstates in Sec.~\ref{Sec:PreludeGauge}, as well as to the orientability of the real part of the eigenvectors in Sec.~\ref{Sec:PreludeVector}. 

\subsection{Topological classifications}

\subsubsection{Gauge obstruction and Homotopy Group}\label{Sec:PreludeGauge}

A topological phase occurs in a non-interacting Hamiltonian when the unitary matrix $U$ of its eigenstates lives on a non-trivial hypersphere $\mathbb{S}^{d}$ of $d$ spatial dimension.\cite{foot2} A non-trivial hypersphere implies that there is an obstruction to (global) gauge fixing of the eigenstates in the entire hypersphere.\cite{Nakahara,Fujikawa,Bernevigbook} One requires multiple (say, two) gauge fixings in different parts of $\mathbb{S}^d$ (say, in the north and south hemispheres $\mathbb{S}^d_{\rm N/S}$). The difference between the two gauges gives the Berry connection (a non-removable, singular gauge). Living on $\mathbb{S}^d$, the Berry connection is a pure gauge, and the corresponding gauge group belongs to either $U(N)$ (Abelian) or $SU(N)$ (non-Abelian) group. Hence, every point on $\mathbb{S}^{d}$ has a 1:1 mapping to every point on the space defined by the generators of the corresponding gauge group (which is also a hypersphere $\mathbb{S}^n$ since the gauge group is unitary, where $n$ spans over the number of generators of the gauge group). The identification of the gauge group, therefore, helps mapping the eigenspace to a corresponding $n^{\rm th}$ homotopy group $\pi_{d}(\mathbb{S}^{n}) \cong \mcz$, or  \ZZ, or 0, where $\mcz \in $ integers, $\mczz\in [0,1)$, and 0 denotes a trivial phase. The subscript $d$ in the homotopy group gives the dimension of the parameter space (momentum here), which is also compact $\mathbb{S}^d$ or a torus $\mathbb{T}^d$. This group identification enumerates distinct topological invariants $-$ the winding numbers, in general $-$ for a given system. \cite{AZ,Schnyder,KitaevPerTab,TenfoldRMP,TenfoldOthers,RShankar}  

\subsubsection{Vector bundle and K-group}\label{Sec:PreludeVector}
The above description of the gauge obstruction arises in complex eigenvectors. One may wonder what happens for real eigenvectors, which may arise in real Hamiltonians, in Majorana representations, and/or when certain \PT~ or equivalent symmetric cases \cite{PTEuler,C2T1,C2T}. It turns out that there is a different topological class owing to the obstruction to smooth orientation of the real eigenvectors in the entire BZ. To intuitively bridge its connection to the gauge obstruction, let us start with a complex eigenvector, and locally impose it's phase part to be a constant at all $k$-points. Then the resultant real eigenvectors have an emergent `local' symmetry, say $SO(N)$, and the corresponding eigenvector bundle lives on the tangent space of a compact manifold. Now, as we continuously vary the base point of an eigenvector across the compact manifold and return back to its starting point, the eigenvector will acquire an $n\in\mcz$ - fold rotation to its starting orientation. So $n$ measures the obstruction to contractibility of the collection of vector bundles, and leads to an invariant $-$ called the K-theory (for real vector bundles) or $\tilde{K}$-theory (for complex eigenvectors).\cite{footexample}

The topological phases of matter has been classified within the K-theory by identifying the homotopy class $\pi_d(KO(N))$ in both ten-fold class as well as for various topological crystalline insulators.\cite{Schnyder,KitaevPerTab,TCIClass} In fact, it can be proved that both the winding numbers for real vector (defined by obstruction to orientability) and for the corresponding complex eigenvectors (defined by obstruction to gauge fixing) are the same when both invariants are defined.\cite{PTEuler,C2T1}

In the present context of symmetry breaking topological phases, one may wonder if any of the symmetry breaking example corresponds to a migration from the gauge obstruction class to the K-theory class. It is however a possibility, especially for certain \PT-invariant cases as well as in some topological materials protected by crystalline symmetries.\cite{TCIClass,PTEuler,C2T1,C2T} Although, we have several examples, where the broken anti-unitary \T~ or \C~ symmetries are restored by antiunitary \PT~ or \CP~ or unitary \CT~ or \CPT~ combinations, but the eigenvectors remain complex, and acquire the same gauge obstruction between the two regions of the BZ which are related by the original broken \T or \C~ symmetries. Of course, a complex $\tilde{K}$-theory cannot be ruled out, but the essence of our work is that the symmetry breakings do not change the topological invariant and the same topological invariants survive to the broken symmetry phases.


\subsubsection{Internal symmetries and Hilbert space dimensions}
Internal symmetries of the eigenvectors play an important role to the topological classification. When the eigenvectors have an $SU(N)$ symmetry (or $SO(N)$ for real eigenvectors), the corresponding Berry connection is also valued in the $SU(N)$ Lie Algebra. In that case, the topological space is identified by the homotopy group $\pi_d(SU(N))$, and the corresponding topological invariant depends on $N$. Most of the topological classifications are performed in the infinite limit of $N$.

In principle, the internal symmetry structure and the dimension of the Hilbert space are automatically incorporated in the classification scheme. Because the number of generators for a given group $SU(N)$ dictates the eigenvectors to live in a compact target space $S^n$. Then the mapping to the parameter space $\mathbb{S}^d$ gives the homotopy group $\pi_d(S^n)$ as discussed above. However, differences arise when the representation is reducible in that there is an irreducible representation in which the homotopy classification changes. A simple example would be $\pi_2(SU(N))\cong\mczz$ for $N>2$, and \Z~ otherwise. In what follows, if we have a four-component spinor represented by $SU(2)\times SU(2)$, then each $SU(2)$ block gives a topological invariant in \Z~ class, but with opposite sign so that the net invariant is defined in the \ZZ~ class.



\subsubsection{Role of discrete symmetries}
So far, no discrete symmetry is implemented. It turns out that when discrete symmetries (such as TR, CC, chiral,  parity/inversion, and/or others) are present, the unitary matrix $U$ of the complex eigenvector (or orthogonal matrix for real eigenvectors) can be brought down to follow a lower quotient  group. (The same dissection of the eigenvector space to quotient group can also be done with discrete crystalline (point, space or non-symmorphic) group symmetries, giving the so-called topological crystalline phases classifications.) Depending on whether one or more discrete symmetry (s) is present, the quotient group can be uniquely defined. Hence the corresponding relation between the $n^{\rm th}$ homotopy class, or Euler/Stefer-Withney class for vector bundle, and the hyperspace dimension $d$ can uniquely dictate its topological group to be either \Z~ or \ZZ~ or 0.\cite{AZ,Schnyder,KitaevPerTab,TenfoldRMP,TenfoldOthers} This result is supposed to be independent of the form of $U$, and hence of the Hamiltonian, and depends only on the symmetries.

\subsection{Symmetry Anomaly and Topological Invariants}
Let us give an alternative and simplified view on how the discrete symmetry can be used to remedy the obstruction, and how it thereby becomes \textit{anomalous}. Let there be two eigenstates $\psi_{\pm}$ to be the symmetry partners for a discrete symmetry $\mathcal{O}$. They can be either (i) the Kramer's partners or particle-hole partners $\psi_+$ and $\psi_-=\mathcal{O}\psi_+$ for antiunitary operator $\mathcal{O}=\mct$, \C$\:$ symmetries, respectively; or, (ii) $\psi_{\pm}$ are two eigenstates: $\mathcal{O}\psi_{\pm}=\pm \psi_{\pm}$ of a unitary operator $\mathcal{O}$ = chiral, parity operators. Suppose now there exists a `band inversion' between these two states across a characteristic momentum ${\bf k}_0$. Then, in the non-trivial phase, one cannot define a smooth gauge at all momenta in the BZ. 
However, one can separately fix the gauge as follows. One gauge for all ${\bf k}<{\bf k}_0$, and another in the ${\bf k}>{\bf k}_0$.  The gauge difference between them is the Berry connection. Without loosing generality, we can claim $\psi_{\pm}$  to live on different gauge-inequivalent regions of the $\mathbb{S}_{\rm N/S}^{d}$. Equivalently, we can say, for a given band (say, valence band), it's eigenvector in the first region (${\bf k}<{\bf k}_0$) belongs to $\mathbb{S}_{\rm N}^{d}$, say, and that in the remaining region belongs to $\mathbf{S}_{\rm S}^{d}$. In other words, the symmetry partners can be used to index different gauge-inequivalent regions of the hypersphere, i.e., $\mathbb{S}^{d}_{\rm N/S}\equiv\mathbb{S}^{d}_{\pm}$. (This description is equivalent to the identification of the quotient group we discussed above.)

According to 't Hooft anomaly analysis,\cite{tHooft} to remedy the gauge obstruction in a non-trivial topological phase, one requires to violate some other symmetry $-$ hence the corresponding symmetry is termed as \textit{anomalous}.  This is understood as follows. Each time,  one moves across $\mathbb{S}^{d}_{\rm N/S}$ ($\equiv\mathbb{S}^{d}_{\pm}$ as distinguished now by the symmetry partners), a singular gauge (Berry connection) has to be removed/added. Since the global symmetry is connected to charge conservation, a `topological charge' has to (one-way) flow across the symmetry partners. This phenomenon is called \textit{anomaly inflow}. In the case of a chiral (parity) symmetry, different chiral (parity) eigenstates distinguish the two gauge inequivalent regions. As a `charge' flows from one chiral (parity) mode to another, we obtain the famous chiral (parity) anomaly.\cite{ABJAnomaly,ThoulessPump,HaldaneQH,ParityRedlich} One of the anomaly indicator of this phase in 2+1 dimension is the first Chern number. Similarly, for the TR case, Kane-Mele proposed a pumping of TR polarization in which an anomaly inflow occurs between the two Kramer's partners. This can be identified as the TR anomaly (also known as spin or helicity pumping in different contexts).\cite{KaneMele,SCZQSH} Similarly, splitting of Majorana pairs to different edges is a representative example of the  charge conjugation anomaly\cite{KitaevMajorana}. Below, we revisit the key topological invariants for these polarizations.  

\subsubsection{TR polarization}\label{Sec:TR}
The TR polarization formula was deduced by Kane and Mele (KM).\cite{KaneMele} We briefly review it in a slightly different way to connect it to the gauge fixing procedure introduced before. We split the Hilbert space for the two Kramer's pair $\psi_{\pm}({\bf k})$ living on $\mathbb{S}_{N/S}^2\equiv \mathbb{S}_{\pm}^2$. Since TR gives ${\bf k}\rightarrow -{\bf k}$, we can also split the torus $\mathbb{T}^2$ into $\mathbb{T}_{\pm}^{2}$ for ${\bf k}> 0$, and ${\bf k}< 0$, respectively. The two Kramer's partners differ by a non-trivial (also called large) gauge rotation $\Omega_{\bf k}$ as $|\psi_+(-{\bf k})\rangle=\Omega_{\bf k}|\mct \psi_-({\bf k})\rangle$.\cite{footLargeGauge} Then the Berry connections for the two Kramer's partners ($\mathcal{A}^{\pm}_{{\bf k}}$) are related to each other by a pure gauge transformation, defined as $\mathcal{A}^+_{-{\bf k}}=\mathcal{A}^-_{{\bf k}} - i\Omega_{\bf k}^{-1}\nabla \Omega_{\bf k}$.\cite{FootGaugefixing,Vanderbilt} If we choose a closed path on the boundary of the BZ ($\partial \mathbb{T}^2$), according to fibre bundle theory, the gauge transformation $\Omega_{\bf k}$ defines a mapping $\partial \mathbb{T}^2 \rightarrow \mathbb{S}^1$. In this case, the gauge transformation has a winding number. However, due to the TR symmetry, the total wraping over $\partial \mathbb{ T}^2$ consists of two wrapings: one clockwise wraping around $\partial \mathbb{T}^2_+$ , followed by an anti-clockwise wraping around $\partial \mathbb{T}^2_-$, yielding a null net winding number. However, if we restrict the loop within $\partial \mathbb{T}^2_+$ we obtain a topological index  $\nu$ of the gauge transformation $\Omega_{\bf k}$ defined as $\nu=\frac{1}{2\pi}\oint_{\partial {\bf T}^2_+} d{\bf k}\cdot {\rm Tr}\left[\Omega_{\bf k}^{-1}\nabla \Omega_{\bf k}\right]$. According to the Atiyah-Singer index theory,\cite{AtiyahSinger} the net anomaly inflow between the Kramer's partners is equal to $\nu$. Since the anomaly inflow leads to an imbalance in the occupancy of the two Kramer's partners, we refer to this process as the TR polarization $-$ a TR anomaly. This is how the topological gauge obstruction is remedied by a TR anomaly.

 With further mathematical treatments, Fu, Kane, and Mele deduced a working formula for $\nu$ as\cite{FuKaneMele}:
\begin{eqnarray}
(-1)^{\nu} = \prod_{n=1}^N\delta_{{\bf k}_n},
\label{Eq:Z2inv}
\end{eqnarray}
where $n$ indices the number of TR invariant ${\bf k}_n$-points in the first quadrant of the BZ. Here $\delta_{\bf k}=\sqrt{{\rm det} (\omega_{\bf k})}/{\rm Pf (\omega_{\bf k}})$, where $\omega_{\bf k}$ is the anti-symmetric sewing matrix with two off-diagonal terms containing $\pm \Omega_{\pm\bf k}^{-1}\nabla \Omega_{\pm\bf k}$. Since $\omega_{\bf k}$ is an anti-symmetric matrix, it's Pfaffian follows  ${\rm Pf (\omega_{\bf k}})=\pm  \sqrt{{\rm det} (\omega_{\bf k})}$. Hence, $\delta_{{\bf k}_n}=\pm 1$, taking  negative sign when an  anomaly inflow (or band inversion) occurs between the two Kramer's partners.  Therefore, according to Eq.~\eqref{Eq:Z2inv}, $\nu=1$ if there is an odd number of anomaly inflow (s) or band inversions between $\psi_{\pm}$; otherwise $\nu=0$. This is how the TR symmetry restricts the winding number $\nu$ belonging to a \ZZ~group.

\subsubsection{Particle-hole polarization}\label{Sec:PH}

By particle-hole symmetry, we mean here that there exists an operator (unitary or antiunitary) $\mathcal{O}$ which anticommutes with the Hamiltonian $H$ such that the energy eigenvalues come in pair $\pm |E|$. Such a case may arise from either \C$\:$ or \CT$\:$ or \CP$\:$ or \CPT$\:$symmetries. Depending on the symmetry, the particle-hole pairs correspond to states at ${\bf k}$ and $\bar{{\bf k}}=\pm {\bf k}$, see Table~\ref{Tab:Sym}. The particle-hole symmetric Hamiltonian can be block-off-diagonalized by a similarity transformation, and we express it as $H=\begin{pmatrix}
	{\bf 0}  &  Q_{\bf k}  \\
	Q^{\dag}_{\bf k}  &  {\bf 0}\\
\end{pmatrix}$. Then the Hilbert space is also split into $\Psi_{\bf k}=(\psi_+({\bf k})~~ \psi_-(\bar{\bf k}))^{T}$, where the particle-hole conjugates are related to each other by $\psi_+({\bf k})=Q_{\bf k}\psi_-(\bar{{\bf k}})$, and $\psi_-(\bar{\bf k})=Q^{\dag}_{\bf k}\psi_+({\bf k})$. This implies that $Q_{\bf k}$ is unitary and is qualitatively equivalent to the (large) gauge transformation $\Omega_{\bf k}$ introduced for the Kramer's partners above. Indeed, explicit Berry connection formulas for the two particle-hole partners reveal that they differ by $\mathcal{A}^+_{\bf k}-\mathcal{A}^-_{\bar{\bf k}}= iQ_{\bf k}^{-1}\nabla Q_{\bf k}$. Following the same argument as done above for the TR case, the anomaly inflow between the two particle-hole partners can be indicated by a winding number $\nu_{\rm ph}=\frac{1}{2\pi}\oint_{\partial {\bf T}^2_+} d{\bf k}\cdot {\rm Tr}\left[Q_{\bf k}^{-1}\nabla Q_{\bf k}\right]$. We notice that $Q_{\bf k}$ may not be Hermitian and hence the winding number can be complex, in general.\cite{GhatakDas}

\subsubsection{Parity polarization}\label{Sec:Parity}

When the system is also invariant under parity, the above description can be cast into an anomaly inflow between different parity states $-$ hence violating parity conservation. We can repeat the above analysis for $\psi_{\pm}({\bf k})$, where now $\pm$ corresponds to two parity eigenvalues. We denote the parity eigenvalue at a given ${\bf k}$ point by $p_{\bf k}$. Fu and Kane showed that the same Eq.~\eqref{Eq:Z2inv} for the winding number $\nu$  also works for parity inflow where $p_{\bf k}$ for the valence band replaces $\delta_{{\bf k}}$.\cite{FuKane} 

\subsubsection{Crystal symmetries}
Because of the contractibility of topological spaces,  the splitting of the BZ into distinct gauge-equivalent subspaces is not unique. As mentioned in the introduction, the symmetry anomaly is a choice especially when there are multiple symmetries. In addition to above discrete symmetries, there are also various discrete crystalline symmetries belonging to point-, space-groups, or non-symorphic groups which can dissect the BZ into different gauge-equivalent space as before. Such topological classifications are collectively called topological crystalline insulators.\cite{TCIClass,TIClassReflection,TIClassSpace}

Is there any crystalline symmetry which can be an alternative choice to provide quantum anomaly in the present symmetry-broken topological phases? While this is generally possible, but it does not apply in our case. Because none of the studied quantum orders break any crystalline symmetry. In various examples in the literature, the broken \T~ or \C~ symmetry combines with a broken to crystalline symmetry to become the defining symmetry anomaly (e.g., transnational symmetry breaking AF phase, or \PT-symmetric topological phase etc).\cite{C2T1,C2T,PTEuler} In the present work, all the order parameters are momentum-independent and do not break any transnational, rotational, or mirror symmetry. Hence the combination of any crystalline symmetry with the broken \T~ or \C~ symmetries remain broken in the ordered phases. In fact, we shall demonstrate that the topological invariants can as well be defined in the continuum limit of the models having continuous transitional and rotational symmetries.

\subsubsection{Chern number}\label{Sec:Chern}


The Chern number is the flux of the Berry curvature $\mathcal{F}_{\bf k}^{\pm}=\nabla\times \mathcal{A}_{\bf k}^{\pm}$ for $\psi_{\pm}$ states in a torus $\mathbb{T}^2$. So, Chern number measures the flux flow between $\psi_{\pm}$ states. In our above description, this amounts to a flux flow between the two hemispheres $\mathbb{S}^2_{\rm N/S}\equiv \mathbb{S}^2_{\pm}\equiv\mathbb{T}^2_{\pm}$. Now if $\psi_{\pm}$  are identified as the symmetry partners of a discrete symmetry, then the flux flow gives the corresponding symmetry anomaly flow.  For example, the Thouless charge pump in a Chern insulator is a manifestation of the chiral anomaly. Similarly, the spin,  mirror, valley Chern numbers as deduced in different contexts are associated with anomalies in spin, mirror, and valley symmetries, respectively.  

For the TR invariant case, we identify $\psi_{\pm}$ as the Kramer's partners, in Sec.~\ref{Sec:TR}. Then the Berry flux of each state is $C_{\pm}=\frac{1}{2\pi}\int_{\mathbb{T}^2} \mathcal{F}^{\pm}_{\bf k}d^2{\bf k}$, with $C_{+}=-C_-$ due to the TR symmetry. However, it is easy to see that the difference between the two Chern numbers is related to the gauge transformation $\Omega_{\bf k}$ as $C_s$ = $\frac{1}{2}|C_+-C_-|$=$\frac{1}{4\pi}\int_{\mathbb{T}^2} {\rm Tr}\left[\nabla\times \Omega_{\bf k}^{-1}\nabla{\Omega}_{\bf k}\right]d^2{\bf k}$ $=\nu$.  When spin is a good quantum number, $\psi_{\pm}$ states are nothing but the $\uparrow$, $\downarrow$ spin states, and hence $C_s$ is often called  the spin Chern number. In what follows,  if we want to evaluate individual values of $C_{\pm}$, the Hamiltonian has to be block diagonal with each block breaking the TR symmetry (the total TR is preserved),  and $\psi_{\pm}$ are the eigenvectors of the two block. In such a case, the anomaly flow between the two block Hamiltonians may or may not have any physical interpretation unless there is a physical symmetry (such as spin-rotation for spin Chern insulator) that relates the two block diagonals. 

We can easily generalize the above analysis to the case of the particle-hole partners $\psi_{\pm}$, as discussed in Sec.~\ref{Sec:PH} In this case, the above formalism remains intact as we replace $\Omega_{\bf k}\rightarrow Q_{\bf k}$, the block-off-diagonal part of the Hamiltonian. Thanks to the Nielsen–Ninomiya's {\it no-go} theory,\cite{NielsenNinomiya} the total Chern number between the particle-hole partners must vanish and hence $C_+=-C_-$. However, we notice here that in an insulator $\psi_{\pm}$ states denote the valence and conduction bands. The Hall effect is only contributed by the valence band. So we get a finite Hall effect, and the system is called the quantum anomalous Hall (QAH) insulator or the Chern insulator. Chiral symmetry (\CT) is a prominent example of the particle-hole symmetry and hence the Chern or QAH insulators are often considered as a result of chiral anomaly.

\section{Symmetries to be considered}\label{Sec:DisSym}

\begin{table*}[t]
\renewcommand{\arraystretch}{1.5}
\centering
\begin{tabular}{|m{4cm}|m{6.5cm}|m{1cm}|m{1.5cm}|}
\hline
Operator ($\mathcal{O}$) & Operation & $\mathcal{O}^2$ & Unitary\\
\hline
$\mct=i\g_{13}\mck$  & $(\mct)^{-1} H({\bf k})\mct=H(-{\bf k})$ & $-1$ & no \\
\hline
$\mcc=\g_1\cc$  & $(\mcc)^{-1} H({\bf k})\mcc=-H(-{\bf k})$ & $+1$ & no \\
\hline
$\mcct=\g_3$  & $(\mcct)^{-1} H({\bf k})\mcct=-H({\bf k})$ & $+1$ & yes \\
\hline
$\mcp=\g_4$  & $(\mcp)^{-1} H({\bf k})\mcp=H(-{\bf k})$ & $+1$ & yes \\
\hline
$\mcp\mct=i\g_{25}\cc$  & $(\mcp\mct)^{-1}H({\bf k})\mcp\mct=H({\bf k})$ & $-1$ & no \\
\hline
$\mcc\mcp=i\g_{14}\cc$  & $(\mcc\mcp)^{-1}H({\bf k})\mcc\mcp=-H({\bf k})$ & $-1$ & no \\
\hline
$\mcc\mcp\mct=i\g_{34}$  & $(\mcc\mcp\mct)^{-1}H({\bf k})\mcc\mcp\mct=-H({-\bf k})$ & $-1$ & yes \\
\hline
\end{tabular}
\caption{Various symmetry operators considered here, and their operations on a Bloch Hamiltonian.}
\label{Tab:Sym}
\end{table*}

Here we focus on the ten-fold classification scheme which is based on TR \T,  CC \C, and their product \CT, namely chiral/sublattice symmetries. In addition, we also consider parity \P, and all their dual and trinal combinations. We  refer them as \PT, \CP, and \CPT symmetries.  All the symmetry properties are listed in Table~\ref{Tab:Sym}. 

The parity symmetry can be taken as a place-holder to a wider range of discrete, or even continuous symmetries, if present, which may appear arise to rescue the spontaneous loss of the other symmetries. We refrain from such a wider generalization by focusing on the order parameters which do not  break any space, point-group, translation or non-symmorphic symmetries. Some of the order parameters break parity and thus we focus only on a parity symmetry, defined by spatial inversion and sublattice exchange: $\mcp\psi_{\rm A,B}({\bf k}) =\psi_{\rm B,A}(-{\bf k})$, where A, B stand for sublattice indices. 

An anti-unitary operator $\mathcal{O}$, with $\mathcal{O}^2=-1$, which commutes with the Hamiltonian gives a Kramer's degeneracy. For spin-1/2 Hamiltonian, $\mct$ and $\mcpt$ symmetries satisfy this criterion. For $\mct$, the Kramer's degeneracy occurs only at the TR invariant ${\bf k}$-points. In this case, edge states in a \ZZ~ topological phase produce a gapless Dirac cone at a \T-invariant ${\bf k}$-point. The presence of the $\mcpt$ symmetry gives a two-fold degeneracy at all ${\bf k}$ points.

For our spin-1/2 case without superconductivity, the antiunitary CC operator follows $\mcc^2=+1$. But since $\mcc$ anticommutes with $\mcp$ operator, we obtain $(\mccp)^2=-1$ and $(\mccpt)^2=-1$.\cite{footCPT} This case is interesting and requires a special attention. $\mccp$ is an antiunitary operator, but since it anticommutes with the Hamiltonian, one obtains particle-hole symmetric energy eigenvalues.\cite{foot_PH} However, does a gapless point ensures a Kramer's degeneracy? To find out we consider the inner product of an eigenstate $\psi_{n}({\bf k})$: $\langle {\psi_{n}({\bf k})}|{\mccp\psi_{n}({\bf k})}\rangle=\langle{\mccp\psi_{n}({\bf k})}|{(\mccp)^2\psi_{n}({\bf k})}\rangle^*=-\langle{\psi_{n}({\bf k})}|{\mccp\psi_{n}({\bf k})}\rangle=0$ at every ${\bf k}$-points. In the first step, we use the antiunitary property, and in the second step we substitute $(\mccp)^2=-1$.  Therefore, the gapless points at any ${\bf k}$ point are two-fold degenerate. 

The sublattice symmetry is unitary and $(\mcct)^2=+1$.  For two eigenenergies $\pm E_n$, the corresponding eigenstates $\psi_{n}({\bf k})$ and $\mcct \psi_{n}({\bf k})$ belong to the same Hilbert space, and hence are orthogonal. Interestingly, $(\mccpt)^2=-1$, but its a unitary symmetry. However, unlike the \CP~case discussed above, \CPT$\:$ symmetry guarantees no degeneracy in the Hilbert space. The  $\mccpt$ invariance implies $\mccpt H({\bf k})=-H({-{\bf k}})\mccpt$. Hence $H({\bf -k})\mccpt\psi_{n}({\bf k})=-E_{n}(-{\bf k})\mccpt\psi_{n}({\bf k})$. For a parity invariant case $H({\bf k})=H(-{\bf k})$, the \CPT~symmetry becomes analogous to the \CT~case. Here, the energy spectrum is particle-hole symmetric, so $\psi_{n}({\bf k})$, $\mccpt \psi_{n}({\bf k})$ are orthogonal. However, when the parity is absent, and that $E_{n}({\bf k})\ne E_{n}(-{\bf k})$ except at the high-symmetric ${ \bf k}$ points (also known as TR, if present, invariant ${\bf k}$-point). However, since \CPT$\:$ is unitary, such a degeneracy, if exists, is not classified as a Kramer's degeneracy. \CPT$\:$ symmetry is also considered earlier in topological insulators and superconductors.\cite{CPT1,CPT2}

\begin{figure}[ht]
\centering
\includegraphics[scale=0.35]{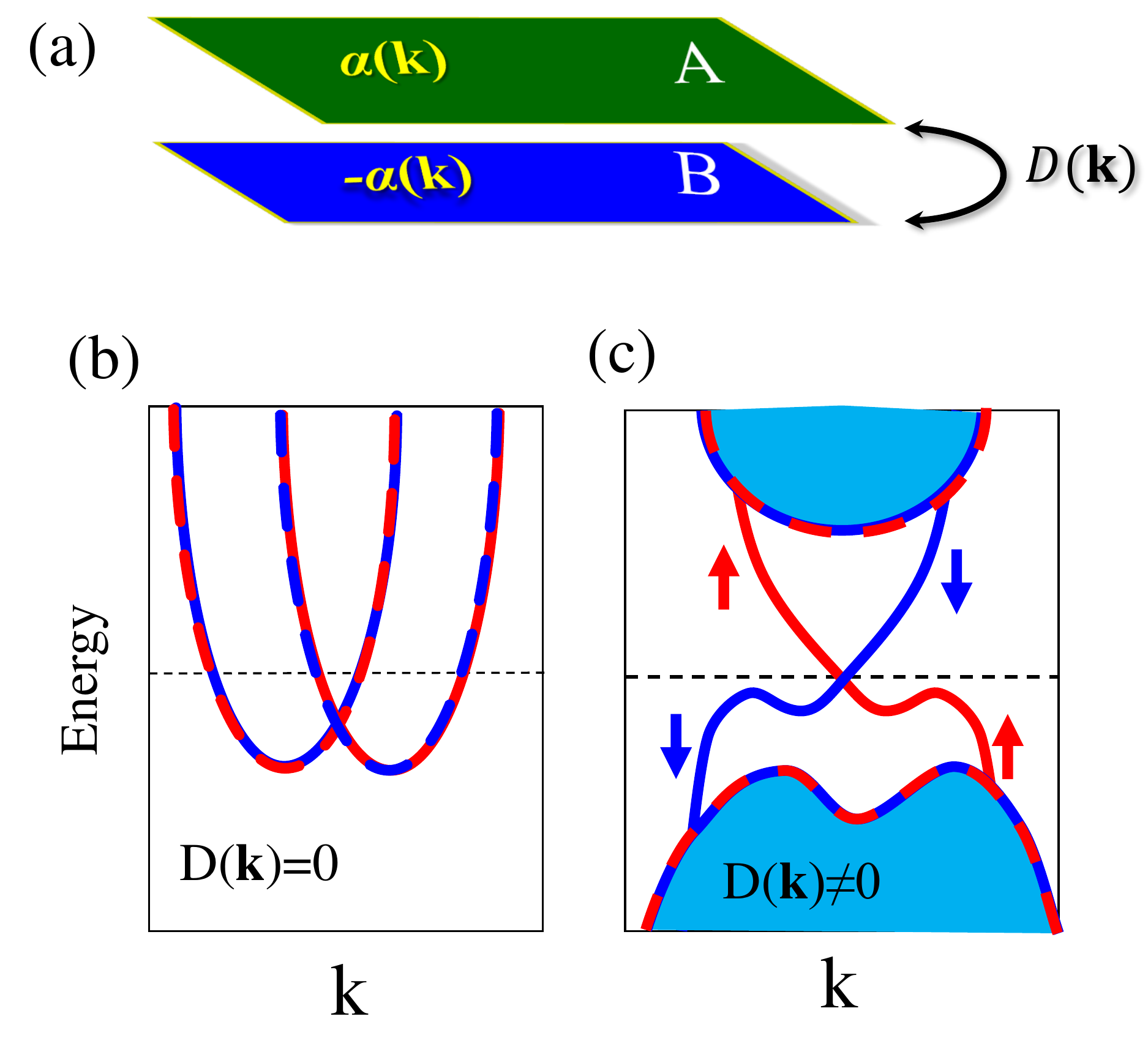}
\caption{(a) Schematic diagram of the Rashba bilayer with opposite Rashba SOC $\pm \alpha({\bf k})$, and connected by spinless coupling $D({\bf k})$. (b) The band dispersion of a uncoupled ($D({\bf k})=0$) Rashba-bilayer, giving a spin-degenerate band structure. Here red and blue color denote up and down spin states. (c) A cartoon band structure of the Rashba-bilayer for finite $D({\bf k})$ with blue shadded regions giving bulk valence and conduction bands, while the blue and red solid lines are the helical edge states.}
\label{Fig:Rashbabilayer}
\end{figure}

\section{The Hamiltonian}

We consider a bilayer of 2D lattice with Rashba-type SOC; but with opposite SOCs in the two adjacent layers, see Fig.~\ref{Fig:Rashbabilayer}. We denote it as the Rashba bilayer. The $2\times 2$ Hamiltonian for each Rashba monolayer in the spin-1/2 basis is written as $h_{{\rm A},{\bf k}} = \xi_{{\bf k}} \sigma_0 + {\bm \alpha}_{\bf k}\cdot {\bm \sigma}$. $\xi_{\bf k}=-2t(\cos{k_x} + \cos{k_y})-\mu$ is the intra-layer dispersion in 2D lattice with $t$ being the nearest-neighbor hopping amplitude, and $\mu$ is the chemical potential. ${\bm \alpha}_{\bf k}$ is the Rashba SOC with components $\alpha_{\bf k}^x=\alpha_{\rm R}\sin{k_y}$, and  $\alpha_{\bf k}^y=\alpha_{\rm R}\sin{k_x}$, and $\alpha_{\rm R}$ is a real constant. $\sigma_{i}$ are the $2\times 2$ Pauli matrices in the spin basis with $\sigma_0$ is the unity matrix. We call it the `A' layer. The adjacent layer, namely the `B' layer, has the same dispersion but with opposite helicity:  $h_{{\rm B},{\bf k}} = \xi_{{\bf k}} \sigma_0 - {\bm \alpha}_{\bf k}\cdot {\bm \sigma}$.

To ensure a band inversion with a topological phase, we consider an anisotropic, non-spin flip tunneling matrix-element between the two layers as 
$D_{\bf k}= D_0 + 2D_1(2-\cos{k_x}-\cos{k_y})$,
where $D_0$ is the onsite and $D_1$ is the nearest neighbor, out-of-plane, hopping coefficients (real). The full Hamiltonian becomes
\begin{eqnarray}
H_{0}=
\begin{pmatrix}
h_{{\rm A},{\bf k}} & D_{\bf k}\mathbb{I}_2\\
D_{\bf k}\mathbb{I}_2 & h_{{\rm B},{\bf k}} \\
\end{pmatrix}= \xi_{\bf k}\mathbb{I}_4 + \Gamma_{1}\alpha_{\bf k}^x+\Gamma_{2}\alpha_{\bf k}^y + \Gamma_{4}D_{\bf k}.
\label{Ham1}
\end{eqnarray}
Here the three ${\Gamma}_i$ matrices are defined as $\Gamma_{1,2,3}=\tau_z\otimes\sigma_{x,y,z}$ and $\Gamma_{4,5}=\tau_{x,y}\otimes \sigma_{0}$, where $\tau_i$ and $\sigma_i$ are the $2\times 2$ Pauli matrices in the sublattice and spin space, respectively, and $\sigma_0$ is a $2\times 2$ unit matrix. A full list of $\Gamma$ matrices are given in Appendix~\ref{AppendixA}. 

The present heterostructure belongs to the $D_{4h}$-group, possessing a four-fold rotational symmetry, and in-plane inversion symmetry, and a mirror symmetry between the two layers. Therefore, the \P~operation consists of ${\bf k}\leftrightarrow -{\bf k}$, and sublattice inversion ${\rm A}\leftrightarrow {\rm B}$. The Hamiltonian respects \T, \P, \C~symmetries, and hence the chiral \CT, \PT, \CP, and \CPT~symmetries. All symmetry transformation of this Hamiltonian are given in Table~\ref{Tab:Sym}  


The Rashba bilayer model in Eq.~\eqref{Ham1} was introduced earlier by one of us for engineering 2D and 3D topological insulators in heterostructure.\cite{DasRashba,DasSODW}  Such a Rashba-bilayer (and its equivalent family of bilayers with opposite SOC) is also shown to possess hidden spin polarization in real space in different layers, while the bands are spin-degenerate in the momentum spare owing to \PT~ invariance.\cite{HiddenSpin}. In non-centrosymmetric layered materials, such as BiTeCl, similar opposite SOC is observed in adjacent layers, which presumably plays a role for its 3D \ZZ~ topological phase.\cite{BiTeCl}

\subsection{Spontaneous symmetry breaking perturbations}
With an eye on finding symmetry breaking order parameters in the above setup, we consider an extended Hubbard model with intra- ($U$) and inter-sublattice ($V$) onsite interactions,

\begin{equation}
H_{\rm int}=U\sum_{\alpha\in(\rm A,B)}n_{\alpha\uparrow}n_{\alpha\downarrow} + V \sum_{\alpha\ne \beta\in(\rm A,B)}n_{\alpha}n_{\beta},
\label{Eq:Hubbard}
\end{equation}
where $\alpha$, $\beta$ denote layer/sublattices indices. $n_{\alpha s}$ is the number operator for the $\alpha^{\rm th}$ sublattice with spin $\uparrow,\downarrow$. $n_{\alpha}=n_{\alpha\uparrow}+n_{\alpha\downarrow}$ is the corresponding total density. The spin-operator for a given sublattice $\alpha$ is defined as usual: ${\bm  S}_{\alpha} =\sum_{s,t\in(\uparrow,\downarrow)}\psi^{\dag}_{\alpha s}{\bm \sigma}_{st}\psi_{\alpha t}$, where ${\bm \sigma}$ are the Pauli matrices in spin basis. In analogy, we define sublattice or pseudospin operators for a given spin `$s$' as ${\bm  T}_{s}=\sum_{\alpha,\beta\in (\rm A,B)}\psi^{\dag}_{\alpha s}{\bm \tau}_{\alpha\beta}\psi_{\beta s}$, where ${\bf \tau}$ are the Pauli matrices defined in the sublattice basis. 

Eq.~\eqref{Eq:Hubbard} can produce various order parameters, however, we focus here on five different order parameters which give distinct topological phases. (1) A sublattice density (SD) order which breaks \C~ and \P~ symmetries, but preserves \T~symmetry, and (2) Four different TR symmetry breaking order parameters which in addition, may or may not break \C, and \P~ symmetries.
We are going to include all the order parameters within the mean-field approximation, so that the the interaction terms appear as additional mass terms to the non-interesting Hamiltonian in Eq.~\eqref{Ham1}, and thereby  the total Hamiltonian breaks certain symmetries but may remain adiabatically connected to the non-interacting Hamiltonian.  

\subsubsection{Charge conjugation symmetry breaking state}
We first consider a SD order parameter 
\begin{equation}
\mathcal{N}^{z}=\frac{1}{2}\Big(\langle T^{z}_{\uparrow}\rangle + \langle T^{z}_{\downarrow}\rangle\Big),
\end{equation}
where the expectation value is taken over the corresponding ground state. The corresponding exchange energy is $E_{\rm SD}=\bar{V}\mathcal{N}^z$, where $\bar{V}$ is the effective mean-field coupling constant which can be evaluated from Eq.~\eqref{Eq:Hubbard}. Therefore, the mean-field perturbation to the non-interaction Hamiltonian $H_0$ can be expressed as
\begin{equation}
H_{\rm SD} = E_{\rm SD} \Gamma_{45},
\label{Eq:Parity}
\end{equation}
where $E_{\rm SD}$ is a real number. $E_{\rm SD}$ gives an onsite energy difference between the two sublattices A and B, and hence it breaks \C~ and \P~symmetries. Since TR symmetry remains intact, this order parameter also breaks the sublattice \CT~symmetry.  

\subsubsection{Time-reversal symmetry breaking states}
Next, we focus on four different TR symmetry breaking mean-field order parameters which give distinct topological phases:
\begin{eqnarray}
\mathcal{M}^{\pm}&=&\frac{1}{2}\Big(\langle S^z_{\rm A}\rangle \pm \langle S^z_{\rm B}\rangle\Big),\\
\mathcal{N}^{x/y}&=&\frac{1}{2}\Big(\langle T^{x/y}_{\uparrow}\rangle \mp \langle T^{x/y}_{\downarrow}\rangle\Big),
\end{eqnarray}
where the expectation value is taken over the corresponding ground states. The first two terms $\mathcal{M}^{\pm}$ are easily identified as ferromagnetic (FM) and antiferromagnetic (AF) orders, respectively. $\mathcal{N}^{x/y}$ are unusual order parameters, giving sublattice magnetic (SM) and sublattice vortex (SV) orders, respectively. For $\mathcal{N}^{x}$, the difference between the two spin components $(T^{x}_{\uparrow,\downarrow})$ helps breaking the TR symmetry.  Among the sublattice operators ${\bf T}_{s}$, only $T^y_{s}$ component breaks TR symmetry (due to the presence of $i$ in $\tau_y$). Hence $\mathcal{N}^{y}$ also breaks TR symmetry (it will become evident in Sec.~\ref{Sec:AFSV} why we name this state a SV state). The magnitude of the resulting exchange energy from all four order parameters can be collectively defined as $E_{\rm FM/AF}=\bar{U}\mathcal{M}^{\pm}$ or $E_{\rm SM/SV}=\bar{V}\mathcal{N}^{x/y}$, where $\bar{U}$, and $\bar{V}$ are the effective mean-field coupling constants. It it clear that AF and FM terms dominate in the limit of $V\rightarrow 0$, while the other two arise in the case of $U\rightarrow 0$. Since our key purpose is not to study the quantum phase transition, rather the topological phase transition induced by these order parameters, we do not discuss further any quantum phase diagram of these order parameters. 

The corresponding mean-field perturbations to the non-interaction Hamiltonian $H_0$ can be expressed in terms of the $\Gamma$-matrices as:
\begin{eqnarray}
H_{\rm FM}&=& E_{\rm FM}\g_{12}, \qquad   H_{\rm AF}=E_{\rm AF}\g_{3},\nonumber\\
H_{\rm SM}&=&-E_{\rm SM}\g_{35},\quad H_{\rm SV}=-E_{\rm SV}\g_{5}.
\label{Eq:Perturbation}
\end{eqnarray}
In all cases, we treat the exchange energy $E_i$ as an adjustable parameter.

\section{Time-reversal invariant topological phases}

\begin{table*}[ht]
\renewcommand{\arraystretch}{1.5}
\centering
\begin{tabular}{|m{0.3cm}|m{2.5cm}|m{0.65cm}|m{0.65cm}|m{0.65cm}|m{0.65cm}|m{0.7cm}|m{0.7cm}|m{0.95cm}|m{0.9cm}|m{0.9cm}|m{1.5cm}|}
\hline
&$H$ & $\mct$ & $\mcc$ & $\mcct$ & $\mcp$ & $\mcp\mct$ & $\mcc\mcp$ & $\mcc\mcp\mct$ & \multicolumn{2}{c|}{\it Ten-fold way} & Our result\\
\hline
\multirow{3}{*}{\rotatebox{90}{}}&$H_0$ & $-1$ & $+1$ & $+1$ & $+1$ & $-1$ & $-1$ &$-1$ & DIII & $\zz$ & $\zz$\\
\hline
{\rotatebox{90}{SD}}&$H_0+H_{\rm SD}$  & $-1$ & $0$ & $0$ & $0$ & $0$ & $-1$ & $-1$ & AII & $\zz$ & $\zz$\\
\hline
\multirow{3}{*}{\rotatebox{90}{FM}}&$H_0+H_{\rm FM}$  & $0$ & $+1$ & $0$ & $+1$ & $0$ & $-1$ &$0$ & D & $\z$ & $\zz$, $\z$\\
&$H_0+H_{\rm FM}+H_{\rm SD}$ & $0$ & $0$ & $0$ & $0$ & $0$ & $-1$ & $0$ & A & $\z$ & $\zz$, $\z$ \\
\hline
\multirow{3}{*}{\rotatebox{90}{SM}}&$H_0+H_{\rm SM}$  & $0$ & $0$ & $1$ & $+1$ & $0$ & $0$ & $-1$ & AIII & $0$ & $\zz$\\
&$H_0+H_{\rm SM}+H_{\rm SD}$  & $0$ & $0$ & $0$ & $0$ & $0$ & $0$ & $-1$ & A & $\z$ & $\zz$\\
\hline
\multirow{3}{*}{\rotatebox{90}{AF}}&$H_0+H_{\rm AF}$  & $0$ & $+1$ & $0$ & $0$ & $-1$ & $0$ & $-1$ & D & $\z$ & $\zz$\\
&$H_0+H_{\rm AF}+H_{\rm SD}$  & $0$ & $0$ & $0$ & $0$ & $0$ & $0$ & $-1$ & A & $\z$ & $\zz$\\
\hline
\multirow{3}{*}{\rotatebox{90}{SV}}&$H_0+H_{\rm SV}$  & $0$ & $0$ & $+1$ & $0$ & $-1$ & $-1$ & $0$ & AIII & $0$ & $\zz$\\
&$H_0+H_{\rm SV}+H_{\rm SD}$  & $0$ & $0$ & $0$ & $0$ & $0$ & $-1$ & $0$ & A & $\z$ & $\zz$\\
\hline
\end{tabular}
\caption{This table encompass all topological non-trivial phases obtained with and without different symmetry breaking perturbations. The value for each symmetry $\mathcal{O}$ is defined as $\mathcal{O}^2=\pm 1$, if present, and 0 if absent. The column for `ten-fold class' list the expected symmetry of topological invariant between \Z and \ZZ homotopy group and zero means topologically trivial phase. The last column gives the group of the topological invariant obtained in this study. The row for FM includes two topological classes, which means for low value of the FM perturbation, we have a \ZZ-class while above a corresponding critical value, the topological phase changes to \Z-class.}
\label{Tab:TI}
\end{table*}

In Table~\ref{Tab:TI}, we list all different topological phases obtained in the present study. Here in addition to the \T, \C, and \CT~symmetries, we also investigate \P,~\PT,~\CP,~and \CPT~symmetries. 
The value for each symmetry operator $\mathcal{O}$ is defined as $\mathcal{O}^2=\pm 1$, if present, and 0 if absent. In the second last column, we list the expected homotopy group classification of the topological invariant based on the value of \T,  \C~ and \CT~symmetries, as deduced in Refs.~\cite{AZ,Schnyder,KitaevPerTab,TenfoldRMP,TenfoldOthers,TCIClass,TIClassReflection,TIClassSpace,TIClassClifford} The final column gives the topological class deduced in the present study. Below, we individually discuss all the phases, and the corresponding indicator for the topological invariant.  

\subsection{Quantum spin Hall state}

\begin{figure}[t]
\centering
\includegraphics[scale=0.22]{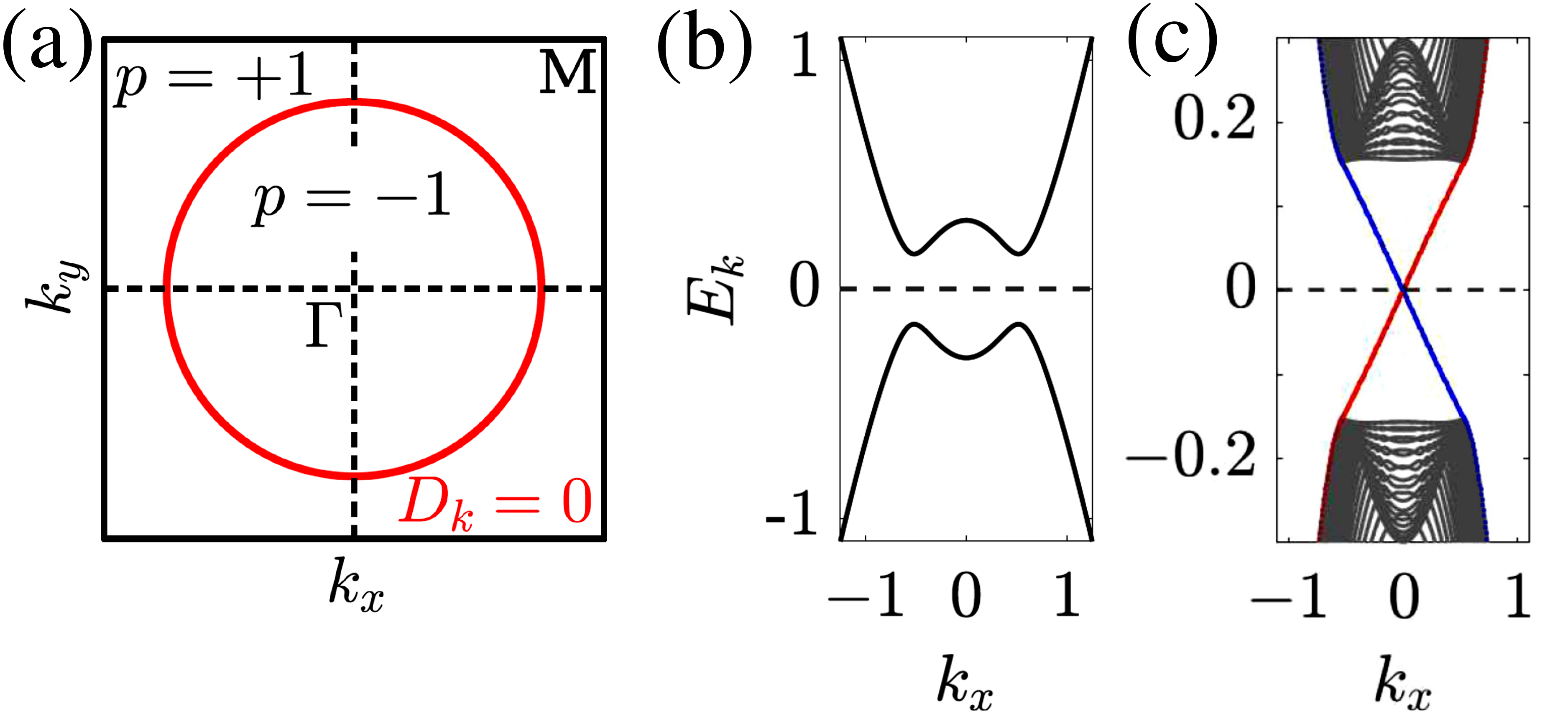}
\caption{(a) $D_{\bf k}=0$ contour in the non-trivial \ZZ~topological phase. Parity ($p$) of an eigenstates changes between $\pm 1$ across this contour. (b) A typical non-trivial bulk band structure is plotted along a high-symmetric direction in the quantum spin-Hall state without any perturbation. Each band is doubly degenerate due to the \PT~invariance. (c) Corresponding edge states with a Dirac cone at the $\Gamma$-point.
}
\label{fig2}
\end{figure}

The non-interacting Hamiltonian $H_0$ in Eq.~\eqref{Ham1} respects all symmetries we are concerned with here. We compute the \ZZ$\:$ topological invariant in terms of the TR symmetry (time-reversal polarization\cite{KaneMele,FuKaneMele}) as well as with parity (parity anomaly)\cite{FuKaneMele}, and Chern number\cite{SCZQSH} wherever applicable. For those systems where the topology is quantified by multiple invariants, i.e., multiple anomalies, we refer them as dual topological insulator. 

\subsubsection{TR and Parity polarization calculations}

We calculate the TR invariant from Eq.~\eqref{Eq:Z2inv} by explicitly calculating the sewing matrix for the TR operator. We find $\nu=1$  in the topological invariant case [see Fig.~\ref{fig2}(b)]. Given that $\Gamma_4$ is the parity operator, the parity eigenvalues are determined by $\delta_{\bf k}={\rm sgn}[D_{\bf k}]$. The result of $\nu$ is also recalculated by using the parity eigenvalues.

Instead of splitting the torus into $\pm {\bf k}$ quadrants of the BZ as often done in the TR case,\cite{FuKaneMele} we split it into two regions $\mathbb{T}^2_{\pm}$ separated by the parity eigenvalues $p_{\bf k}=\pm$. The two parity regions are separated by the boundary made of the $D_{{\bf k}_0}=0$ contour, as shown by red line in Fig.~\ref{fig2}(a). Then, for the valence band, the phase difference between the two regions across the $D_{{\bf k}_0}=0$ boundary gives a non-trivial  Berry connection. In the parameter space of $D_{\bf k}$, as long as off number of nodal contour of $D_{{\bf k}_0}=0$ lie within the BZ, we obtain a non-trivial topological phase. In other words, if there is an odd number of parity inversion in the valence band $-$ equivalently, odd number of charge pumping from one parity state to another $-$ one obtains $\nu=1$, otherwise $\nu=0$. 

\subsubsection{Chern number}

Owing to the \PT invariance, we can also combine the TR and parity anomaly into a \PT-anomaly defined by a corresponding Chern number. 
The \PT invariance guarantees two-fold degeneracy at all ${\bf k}$-points, and hence a block diagonal form of the Hamiltonian exists. The corresponding irreducible representation consists of : $\psi_{b/a,\sigma}=(\psi_{{\rm A}\sigma}\pm \psi_{{\rm B}\sigma})/\sqrt{2}$, at each ${\bf k}$, as the bonding (subscript $b$) and anti-bonding (subscript $a$) states, respectively. For the spinor $\Psi_{\bf k}=(\psi_{b\uparrow}, \psi_{a\downarrow}, \psi_{b\downarrow}, -\psi_{a\uparrow})^{T}$, the non-interacting Hamiltonian in Eq.~\eqref{Ham1} takes the form: 
\begin{eqnarray}
H_0({\bf k}) &=& \xi_{\bf k}\mathbb{I}_4  + 
\begin{pmatrix}
	h^+_{\bf k} &  {\bf 0}    \\
	{\bf 0}   &  h^{-}_{\bf k}\\
\end{pmatrix},\label{BlockH1}\\
{\rm where} && h^{\pm}_{\bf k}=D_{\bf k}\sigma_z \pm\alpha^x_{\bf k}\sigma_x - \alpha^y_{\bf k}\sigma_y.
\label{BlockH2}
\end{eqnarray}
The two eigenvalues for each block are the same: $E^{\pm}_{\bf k}=\xi_{\bf k}\pm\sqrt{|D_{\bf k}|^2+|\alpha_{\bf k}|^2}$ (the eigenstates for the two blocks are sensitive to the sign of coefficient of $\sigma_x$, i.e. $\alpha^x_{\bf k}$, which distinguishes the two blocks with opposite Chern numbers). 

Since each block Hamiltonian $h^{\pm}_{\bf k}$ individually breaks TR symmetry, a Chern number for each block can be defined. We identify $D_{\bf k}$ as the Dirac mass, responsible for the band inversion in both blocks. Interesingly, the parity eigenvalue $p_{\bf k}={\rm sgn}[D_{\bf k}]$ is the same for both block Hamiltonians, and thus both blocks simultaneously possess a topological phase transition. The Chern number in each block is obtained to be  
\begin{equation}
C_{\pm} = \pm \frac{1}{2}\Big[{\rm sgn}[D_{\Gamma}] - {\rm sgn}[D_{M}] \Big],
\label{Chernno}
\end{equation}
where $\Gamma$ and M denote two TR invariant ${\bf k}$-points: $\Gamma =(0,0)$, and ${\rm M}=(\pi,\pi)$, respectively. $\pm$ sign in Eq.~\eqref{Chernno} is deduced from the sign of the coefficient of $\sigma_x$ term in Eq.~\eqref{BlockH2}, i.e. ${\rm sgn}[\alpha^x_{{\bf k}_i}]$. Then the $\mczz$ invariant (defined in Eq.~\eqref{Eq:Z2inv}) is related to the difference between the Chern numbers from the two blocks as $\nu = \frac{1}{2}|C_+-C_-|$. The total Chern number $C_++C_-=0$ for the \PT-invariant cases.

\subsubsection{Analysis}

For this analysis, we set the onsite dispersion $\xi_{\bf k}=0$ without loosing generality. It is easier to go to the continuum limit, and define $\alpha_{\bf k}^{x/y}=\alpha_{\rm R}k_{y/x}$, and $D_{\bf k}=D_0-D_1k^2$, where $k^2=k_x^2+k_y^2$.  (The qualitative analysis remains the same in the tight-binding model in a lattice, where the parameters $\alpha_{\rm R}$, $D_{0,1}$ are renormalized). $D_{\bf k}$ vanishes at the contour of $D_0/D_1=k_0^2$ which gives a constrain that both $D_0$ and $D_1$ must possess the {same sign}. This means $0<D_0/D_1<1$ yields a non-trivial topological phase; otherwise it's a trivial phase. We set $D_1=1$ for convenience of discussion, simplifying the topological condition as $0<D_0<1$. This gives a nodal ring of $D_{\bf k}$ in the $k_x-k_y$ plane centering at the $\Gamma$-point, as shown in Fig.~\ref{fig2}(a). This ensures that the parity ($p_{\bf k}$) at the $\Gamma$-point is reversed from all other \T-invariant ${\bf k}$-point in the BZ, and guarantees a non-trivial, strong topological invariant (according to Eq.~\eqref{Eq:Z2inv} with $\delta_{\bf k}=p_{\bf k}$).

We explicitly calculate the TR polarization using the sewing matrix $\omega_{\bf k}$ in Eq.~\eqref{Eq:Z2inv} and found that $\nu=1$ in the parameter region $0<D_0<1$. The Chern number calculations of Eq.~\eqref{Chernno} also yields $C_{\pm}=\pm 1$ in the same parameter region, otherwise both are zero, as expected. Both the results obtained by TR polarization and Chern number calculations can also be simply understood by the position of the $D_{\bf k}$ nodal ring. Its not captured within the continuum model, but with a lattice model we will see a transition in the Chern number. For small $D_0$, the $D_{\bf k}$ nodal ring encircles the $\Gamma$ points and we get $C_{\pm}=\pm 1$. With further increasing $D_0$ the nodal ring expands, and eventually cross the ($\pi,0$) point to encircles the M-point, the Chern numbers will switch to $C_{\pm}=\mp 1$. In both cases, the TR polarization remains the same as $\nu=1$. Therefore, the parity or helicity polarization/anomaly is reversed with increasing $D_0$, while the TR polarization is insensitive to the direction in which the TR polarization has the \textit{spectral flow}. 

Finally, the emergence of the topological invariant is affirmed by the band structure calculation. For $0<D_0<1$, the bulk band has an inverted band gap, see Fig.~\ref{fig2}(b). The edge state calculation is done by assuming a  finite size lattice of 20 lattice sides along the $y$-direction, and periodic boundary condition along the $x$-direction. We observe a single Dirac cone between the opposite spin-states at the $\Gamma$-point in the topological phase in Fig.~\ref{fig2}(c).   

\subsection{Charge conjugation breaking and parity anomaly}

\begin{figure}[!t]
\centering
\includegraphics[scale=0.48]{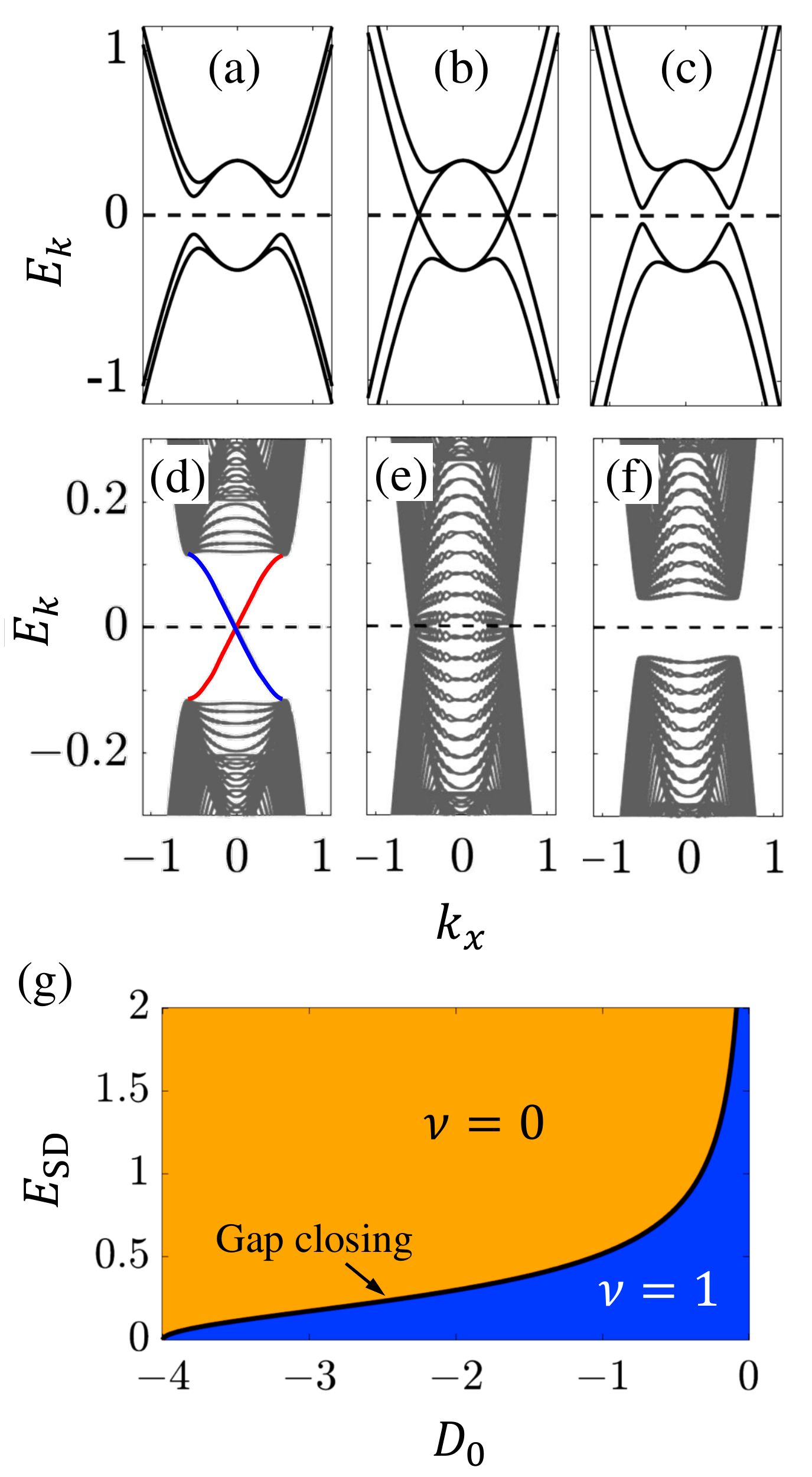}
%
%
\caption{Bulk (top panel) and boundary (middle panel) states for the parity anomaly states defined by $H_0+H_{\rm SD}$. (a) Bulk band structure without two-fold degeneracy except at the TR invariant ${\bf k}-$ points. The bands are, however,   adiabatically connected to the QSH state of $H_0$. (b) Topological critical point where the semimetallic bulk band gap closes for $\eta=-1$ bands in a topological nodal ring, exhibiting an accidental degeneracy in the bulk band structure due to \CP-invariance. (c) Topologically trivial phase. (d-f) Edge states for the non-trivial, critical and trivial phases, respectively. (g) Phase diagram of \ZZ invariant as a function of $D_0$ and $E_{\rm SD}$ (in unit of $D_1=1$) using the lattice model. Blue and yellow regions denote the values of the TR polarization $\nu$ in the non-trivial and trivial phases, respectively, computed using Eq.~\eqref{Eq:Z2inv}.  In this case, Chern number is not defined. The black solid line gives the boundary where the bulk band gap closes (corresponding to (b) case).
}
\label{fig_Parity}
\end{figure}

Our first example demonstrates a topological robustness owing to spontaneous loss of charge conjugation symmetry.  We add the mean-field perturbation of the sublattice density (SD) order (Eq.~\eqref{Eq:Parity} $H_{\rm SD}$ to $H_0$. This perturbation respects \T-symmetry, but breaks \C, and \P, and hence \PT~and \CT~are also broken. Due to the loss of \PT~symmetry, the two-fold degeneracy is lost. However, despite the loss of \C$\:$ and \CT~symmetries, the spectrum shows particle-hole symmetric eigenvalues. This is because the \CP~symmetry is recovered here which provides topological protection to the gapless states at any ${\bf k}$-points, see Fig.~\eqref{fig_Parity}. 

The total Hamiltonian $H_0+H_{\rm SD}$ is no longer block diagonal, and hence the Chern number cannot be defined in this case. Also owing to the loss of parity, the $D_{\bf k}=0$ nodal line cannot be used to separate the two gauge-inequivalent regions. However, TR polarization formula in Eq.~\eqref{Eq:Z2inv} still holds, and gives $\nu=1$, as shown in Fig.~\ref{fig_Parity}(g). Alternatively, the topological invariant can be simply deduced from the adiabatic continuity to the parent Hamiltonian $H_0$. Here the topological phase transition occurs through the gap closing at a two-fold degenerate {\it nodal ring}, rather than a single Dirac point. This can be understood from the energy dispersions: 
\begin{equation}
E^{\pm}_{\eta,{\bf k}}= \pm\sqrt{D_{\bf k}^2+(|\alpha_{\bf k}|+ \eta E_{\rm SD})^2},
\label{Eq:EIG_parity}
\end{equation}
where $\eta=\pm 1$. Given that $E_{\rm SD}$ is a constant number, the gap between the two bands $E^{\pm}_{\eta,{\bf k}}$ does not close for $\eta=+1$. The gap can close for the $\eta=-1$ bands at the $E_F$ at a critical value of $E_{\rm SD}$, if the two conditions $|\alpha_{\bf k_0}| = |E_{\rm SD}|$ and $D_{\bf k_0}=0$ are simultaneously satisfied at ${\bf k_0}\in {\rm BZ}$. Note that $D_{\bf k_0}=0$ is the same nodal ring defined above at $k_0^2=D_0$ (for $D_1=1$). Substituting this in the other condition of $|\alpha_{\bf k_0}| = |E_{\rm SD}|$, we find the critical perturbation energy is $|E_{\rm SD}|=|\alpha_{\rm R}|\sqrt{|D_0|}$. A phase diagram of $\nu$ (color) and band gap (line) as a function $D_0$ and $E_{SD}$ is drawn in Fig.~\ref{fig_Parity}(g). 

For the perturbation energy $|E_{\rm SD}|<|\alpha_{\rm R}|\sqrt{|D_0|}$, we have a topological insulator which is adiabatically connected to the QSH insulator phase at $|E_{\rm SD}|=0$, and thus possess the same \ZZ~topological invariant. However, the symmetry group changes from the DIII to the AII class. Both symmetry groups map to the same \ZZ~ topological group. Interestingly, the topological critical point at $|E_{\rm SD}|=|\alpha_{\rm R}|\sqrt{|D_0|}$ is characteristically distinct from the critical point of the QSH phase. In the present case, the gap closes on a contour, see Fig.~\ref{fig_Parity}(b). This is an accidental two-fold degenerate nodal ring, which is protected by $\mccp$ symmetry. This is in contrast to the topological phase transition of the QSH state at $D_0=0$ or 1, where the band gap closes at a single ${\bf k}=0$ point, and is four-fold degenerate. The latter is a symmetry-enriched Dirac point. 

For the perturbation energy $|E_{\rm SD}|>|\alpha_{\rm R}|\sqrt{|D_0|}$, we have a trivial band insulator, as shown in Fig.~\ref{fig_Parity}(c). From all the results, it is evident that while we consider $|E_{\rm SD}|$ as an energy cost due to the spontaneous development of the SD order parameter, the same conclusion is also valid if the $|E_{\rm SD}|$ term is introduced with an explicit symmetry breaking. This can be obtained by adding a gate voltage between the two layers which hence introduces different onsite potentials to the two sublattice states. 

\section{Time-reversal breaking perturbations }

Next we discuss various TR-breaking perturbations. 

\subsection{FM phase and the survival of \ZZ  invariant }

\begin{figure*}[ht]
\centering
\includegraphics[width=0.9\textwidth]{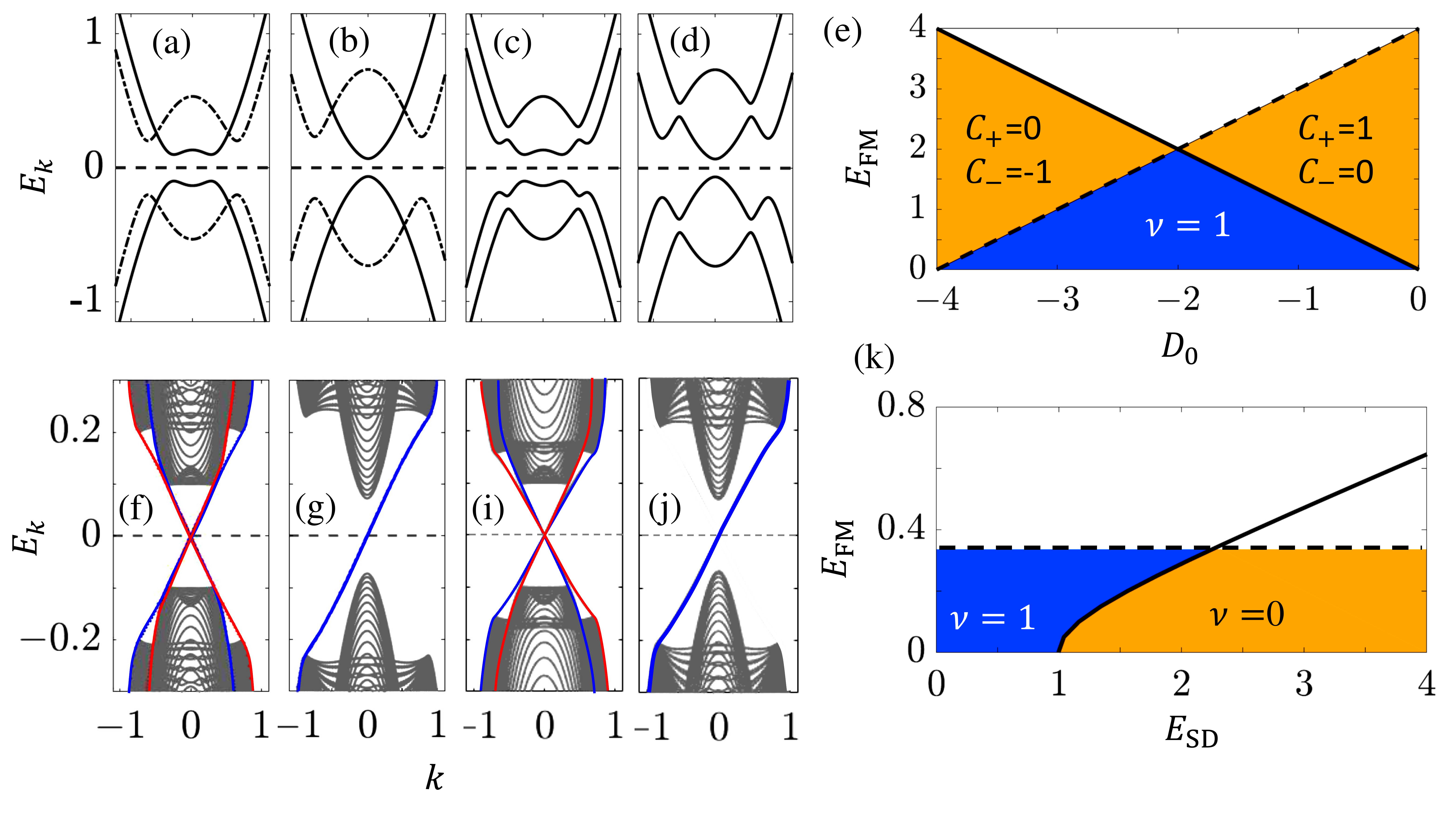}
\caption{(a-b) Bulk band structures of the FM Hamiltonian $H_0+H_{\rm FM}$ are shown in the QSH ($E_{FM}<D_0$) and QAH ($E_{\rm FM}>D_0$) insulator phases, respectively. Here solid and dashed lines are for the bands from two different block Hamiltonians $h_{\bf k}^{\pm}$. (c-d) The same FM band structures but with the parity breaking term included as  $H_0+H_{\rm FM} + H_{\rm SD}$ are, respectively, shown here.  (f-g) Corresponding edge state spectrum for the QSH and QAH insulators exhibit a pair of Dirac cone, and single chiral edge state in each side of the lattice, respectively. (i-j), Edge state for the corresponding top panels, in the presence of \T$\:$ and \P$\:$ breaking phases. (e) Phase diagram of the \ZZ$\:$ invariant $\nu$ and Chern numbers $C_{\pm}$ as a function of $D_0$ and $E_{\rm FM}$. Solid and dashes lines denote the band gap closing in the upper and lower block diagonal Hamiltonians $h_{\bf k}^{\pm}$. Blue region is where TR polarization $\nu$ can be evaluated even if the TR symmetry is lost due to FM order. Orange region is the QAH region, where the Chern number is obtained in either one of the block Hamiltonian, but not in both. Here TR invariant is calculated to be zero. While region is where none of the Chern number is zero, and the TR polarization is undefined. (k) A similar phase diagram is shown for $E_{\rm SD}$ versus $E_{\rm FM}$ in which the TR polarization continues to give $\nu=1$ upto certain value in which the one of bulk band gap is not closed (solid and dashed line). Here Chern number is not defined due to the presence of $E_{\rm SD}$. 
} 
\label{fig_FM}
\end{figure*}

We start with the FM perturbation $H_{\rm FM}$ from Eq.~\eqref{Eq:Perturbation}. $H_{\rm FM}$ breaks \T, while preserves \C, and \P~ symmetries. As a result, \CT~ and \PT~ are broken, and according to the {\it ten-fold way} table, the expected symmetry group belongs to the D class with a topological invariant $\in $\Z~ group (such as a quantum anomalous Hall state with a single chiral edge state). 
In earlier studies, it was found that the FM exchange energy gives a topological phase transition from the QSH state (\ZZ~ class) to the QAH (\Z~class) effect.\cite{QAHReview}

What we find interesting is that the QSH state survives upto a critical value of the exchange energy $E_{\rm FM}$, despite the loss of the \T-symmetry, see Table~\ref{Tab:TI}. Moreover, the TR polarization $\nu=1$ continues to the a valid indicator of the the topological phase even when \T~ is broken. And, the edge states continue to exhibit gapless Dirac cone as in the non-magnetic case. We find that $(\mccp)^2=-1$ is the only symmetry present here, which does not give a band degeneracy unless accidental gapless points occur between the two particle-hole symmetric energy bands. Hence, we find that under the \CP~ symmetry, the band degeneracy inside the bulk is lost, while the gapless Dirac cone at the edge remains intact.
 
The FM perturbation translates into modifying the Dirac mass differently in the two block Hamiltonians in Eq.~\eqref{BlockH1} as
\begin{eqnarray}
&& \quad h^{\pm}_{\bf k}=D_{\bf k}^{\pm}\sigma_z \pm\alpha^x_{\bf k}\sigma_x - \alpha^y_{\bf k}\sigma_y,\\
\label{BlockHFM1}
{\rm where}~~&&~~~ D_{\bf k}^{\pm}=D_{\bf k}\pm E_{\rm FM}\nonumber\\
&&\qquad~~=(D_0\pm E_{\rm FM}) - D_1k^2.
\label{BlockHFM2}
\end{eqnarray}
Since the rest of the Hamiltonian remains the same, we can define the Chern number of the two block Hamiltonians from Eq.~\eqref{Chernno} by simply tracking how the corresponding Dirac masses $D_{\bf k}^{\pm}$ change sign in the BZ. The condition for the band inversion for the two blocks are $0<(D_0\pm E_{\rm FM})/D_1<1$. We start with $0<D_0<1$ (non-trivial QSH phase) and turn on $E_{\rm FM}>0$ (the result is equivalent for $E_{\rm FM}<0$). It is easy to see that for $F_{\rm FM}<D_0$, the band inversion condition is satisfied in both block Hamiltonians, and hence the Chern number in two blocks remains to be  $C_{\pm}=\pm 1$, giving a finite \ZZ~topological invariant $\nu=1$. Remarkably, if we continue to explicitly calculate $\nu$ by the \T- symmetry sewing matrix, surprisingly, we yield the same result of $\nu=1$. Therefore, {\it the system exhibits a QSH behavior with the \ZZ-topological invariant $\nu=1$, despite the loss of the \T-symmetry}. 
We affirm this result by calculating the edge states in this phase, which shows a four-fold degenerate Dirac cone at the Fermi level, see Fig.~\ref{fig_FM}(f). The survival of the QSH phase with FM order is also obtained in the model of graphene\cite{TRBrokenGrapheneFM} . Moreover, the same result is also obtained with the application of external magnetic field\cite{TRBrokenGrapheneB,TRBrokenHgTe}.

However, as $E_{\rm FM}>D_0$, the band inversion criterion in the lower block ($h^{-}_{\bf k}$) is no longer satisfied. Hence, in the lower-block, the Chern number vanishes ($C_-=0$), while the upper block continues to possess $C_{+}=+1$. Therefore, the topological phase changes from the \ZZ~ class (QSH effect) to the \Z~ class (QAH effect). In this case, edge states loose their helical partner in each side, however, remains chiral, and thus a net charge Hall effect is obtained, see Figs.~\ref{fig_FM}(b),(g). In this respect, the $E_{\rm FM}=D_0$ points can be termed as a critical point of symmetry anomaly transition from a helical to chiral anomaly. Therefore, the \ZZ~ invariant of the four-band model results from the  combination of the two underlying \Z~ class two-band Hamiltonians which are connected by the TR symmetries. Even after the loss of the TR symmetry, the two block Hamiltonians continue to exhibit the same symmetry anomaly as long as adiabatic continuity to the symmetry-invariant phase is valid.

Finally, we add the \C~ and \P~ breaking perturbation $H_{\rm SD}$ to the FM Hamiltonian. We find that despite the loss of individual \C- and \P-symmetries, their combination \CP- remains intact. Hence the above conclusions remain valid here. This is evident in the nature of the edge states, see Fig.~\ref{fig_FM}(i). We find that the two Dirac cones from the two edges in the \ZZ-phase do not split here. Above a critical value of $E_{\rm FM}$, band gap closes in the upper block diagonal Hamiltonain and the system turns into a QAH state (see Fig.~\ref{fig_FM}(d),(j)). Due to the presence of $E_{\rm SD}$ the Chern number is no longer defined, yet the explicit calculation of the \T~invariant gives finite value of $\nu$ as shown in Fig.~\ref{fig_FM}(k).

\subsection{Sublattice magnetic (SM) order and \CPT-symmetric topology}

\begin{figure}[!htp]
\centering
\includegraphics[scale=0.47]{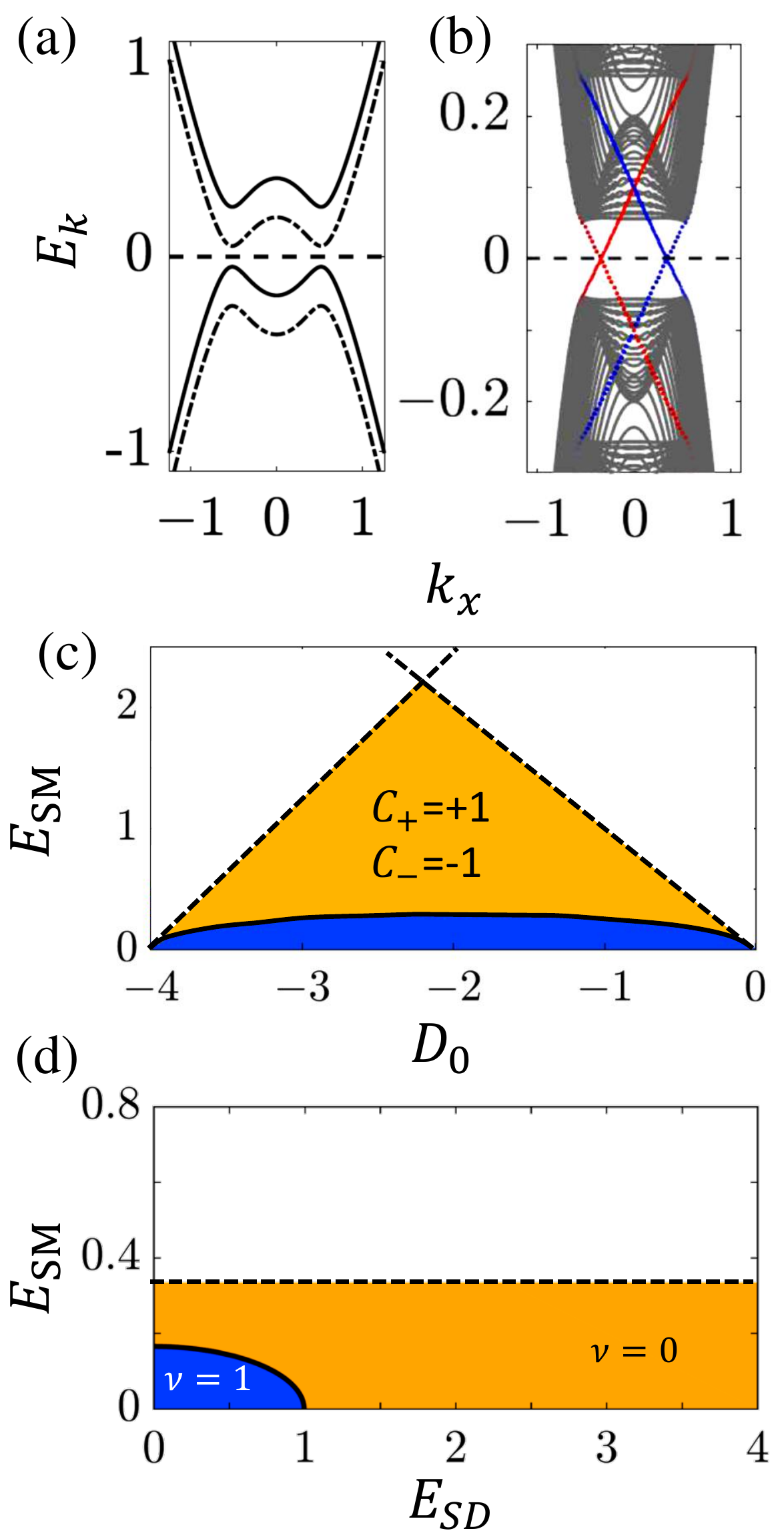}
%
%
%
\caption{(a) Bulk band structure in the SM phase with solid ans dashed lines correspond to eigenvalues for the two blck Hamiltonians. (b) Corresponding edge state profile showing split bands in the momentum axis. (c-d) Phase diagram of topological invariant (as in Fig.~4). For small values of $E_{\rm SM}$ we do find a small region in which the TR polarization continues to give $\nu=1$ despite the loss of TR symmetry. 
}
\label{fig_SM}
\end{figure}

Next we study another puzzling topological phase obtained by the sublattice magnetic (SM) quantum phase. The corresponding perturbation term $H_{\rm SM}$ translates into a simple onsite potential to each block Hamiltonian (Eq.~\eqref{BlockH1}) as:
\begin{eqnarray}
h^{\pm}_{\bf k}=\pm E_{\rm SM} \mathbb{I}_2+ D_{\bf k}\sigma_z \pm\alpha^x_{\bf k}\sigma_x - \alpha^y_{\bf k}\sigma_y.
\label{BlockHCM1}
\end{eqnarray}
The difference between Eq.~\eqref{BlockHCM1} and non-interacting Hamiltonian in Eq.~\eqref{BlockH1} is the overall energy shift by $\pm E_{\rm SM}$. Hence, in addition to the \T-symmetry, the charge conjugation \C~symmetry is also lost here. Their combination \CT~ (sublattice symmetry) remains intact. The system does not possess \PT~and \CP~ symmetries, but the trinal combination of \CPT~ symmetry is invariant here. This, according to the {\it ten-fold way} scheme, should give a \Z~ group, belonging to the AIII cartan class. However, owing to the adiabatic continuity theory of the topological invariance, it is easy to grasp that an irrelevant onsite potential to the topological Hamiltonian does not change its topological invariant, and hence the topological phase remains the same to the non-interacting case such that the two \Z-invariants have opposite sign to give rise to an effective \ZZ~phase. The calculated results of $\nu$ and $C_{\pm}$ are plotted in Figs.~\ref{fig_SM}(c) and (d).

The eigenvalues of Eq.~\eqref{BlockHCM1} are 
\begin{equation}
E^{\pm}_{\eta,{\bf k}}= \pm \Big[E_{\rm SM}+\eta\sqrt{D_{\bf k}^2+|\alpha_{\bf k}|^2}\Big],
\label{Eq:EIG_CM}
\end{equation}
where $\eta=\pm 1$. We show the band diagrams in Fig.~\ref{fig_SM}(a). Solid and dashed curves show the bands from the two different blocks. Each block Hamiltonian is no longer traceless, individually, loosing the particle-hole symmetric spectrum. However, the total Hamiltonian is traceless, implying the presence of a global particle-hole symmetry in the spectrum and that the two blocks act as particle-hole conjugate to each other. This is the {\it \CPT-invariance which guarantees the particle-hole symmetric eigenvalues}.

There are two pairs of edge states in this system, each pair coming from each block Hamiltonian. For the upper block we have two counter-propagating edge modes, possessing a degeneracy at the $\Gamma$-point. These two states are, however, coming from two different edges, as shown in Fig.~\ref{fig_SM}(b), and thus the apparent degenerate point is not a Kramer's degenerate point. The lower block also gives a similar pair of edge modes, shifted below from the other two pairs. Here the two chiral modes have opposite chirality to the ones obtained from the upper block.
 
 More interestingly, at each edge, there are also two helical modes, which are degenerate at a characteristic momentum $\pm {\bf k}^*/2$. This can be interpreted as the the splitting of the Dirac cone into two gapless Weyl excitations in the momentum space. These two modes poses opposite \CPT~eigenvalues, and hence are protected by the Hilbert space orthogonality. This is the emergent  \CPT~invariant topological protection we predict in this work.

The robustness of the \CPT~invariant topological phase becomes more obvious as we include the \P-breaking perturbation $H_{\rm SD}$ to the SM Hamiltonian. As pointed out in Table~\ref{Tab:Sym}, total Hamiltonian $H_{\bf k}=H_0+H_{\rm SM}+H_{\rm SD}$ breaks all the individual symmetries, as well as their dual combinations, and only respects the \CPT~ combination. As shown in Fig.~\ref{fig_SM}(d), the topological phase remains similar upto the critical value of $E_{\rm SD}$. This example gives a novel demonstration of the \CPT~ invariant topological phase in condensed matter systems. 

\subsection{Antiferromagnetic and vortex phases}\label{Sec:AFSV}

\begin{figure}[!htp]
\centering
\includegraphics[scale=0.5]{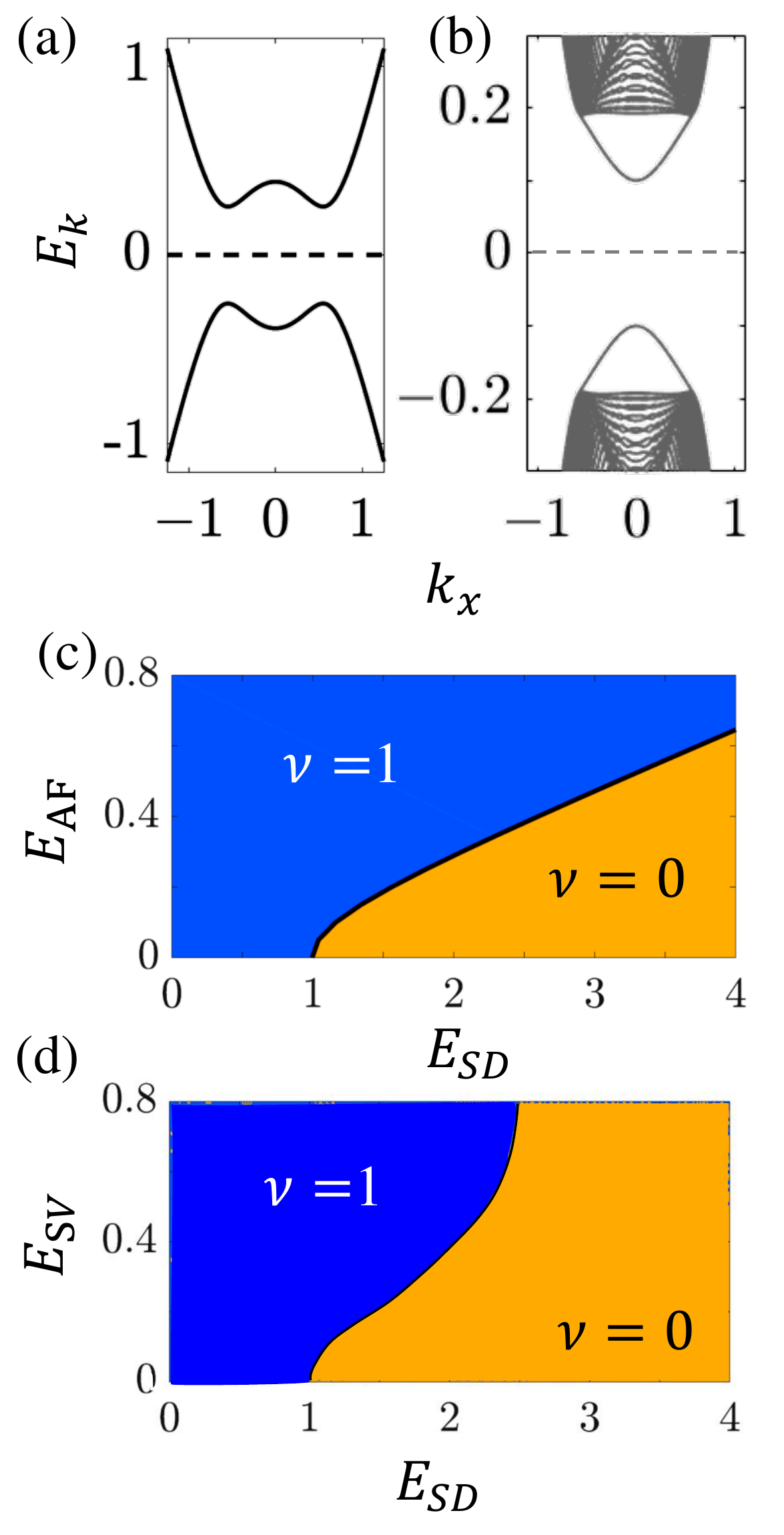}
\caption{(a) Bulk band structure with AF and SV perturbations (both phases give the same band topology). The bands are adiabatically connected to the QSH state without these TR breaking perturbations. (b) The corresponding edge states, showing gapped boundary modes. (c) Phase diagram of TR polarization $\nu$ as a function of $E_{\rm AF}$ and $E_{\rm SD}$. (d) Same as (c) but for the SV phase. As explained in the main text, the TR polarization continues to give $\nu=1$ for all values of $E_{\rm AF}$ and $E_{\rm SV}$, unless there is a gap closing, which is visible for finite value of $E_{\rm SD}$ above which $\nu=0$.
}
\label{fig_AF}
\end{figure}

Next we discuss antiferromagnetic (AF) and sublattice vortex (SV) phases. In both phases, the Hamiltonian breaks both \T~ and \P~, but preserves $(\mcpt)^2=-1$ symmetry. As a result, the bands remain two-fold degenerate at all ${\bf k}-$points, and are adiabatically connected to the non-interacting band structure in {\it both} trivial and non-trivial topological phases. In what follows, the AF and SV phases are topologically connected to the non-interacting Hamiltonian ($H_0$) despite the loss of TR symmetry. However, unlike the FM and SM phases, here the edge states are gapped, as seen in Fig.~\ref{fig_AF}. 

Having the adiabatic continuity preserved, a natural question is: does the system continue to feature the same \ZZ-topological invariant as in $H_0$, or does there arise a new topological invariant? According to the {\it ten-fold~way} table, the AF phase belongs to the \Z~ topological group with the CC symmetry. And the SV phase, without the CC, is a trivial phase. On the contrary, our calculation shows that the \ZZ~ invariant $\nu$ continues to be 1 up to certain parameter values of $E_{\rm AF}$ and $E_{\rm SV}$, see Fig.~\ref{fig_AF}. 

It is important to distinguish first the present case with other \PT~ (or similar discrete space and time) symmetric topological phases discussed in the literature. (1) Both our \PT-invariant AF and SV phases have complex eigenvectors and suffer the same gauge obstruction as in the non-interaction cases. This should be contrasted with the Euler class \PT~invariant (or equivalent discrete space-time symmetric) phases which have real eigenvectors and are classified by the vector bundles within the K-theory.\cite{PTEuler,C2T1,C2T} (2) On the other hand, a 3+1D \PT~invariant AF topological insulator has similar features (gaped bulk and surface states, and adiabatic continuity to the non-magnetic topological insulator), and such an AF state can be described by an axion $\theta$ term\cite{AxionIns}, as observed in recent experiments\cite{MnBiTe}. The axion insulator has two opposite 2+1 D surface states with opposite magnetic polarizarions, and half-integer Hall conductivity. The two surfaces are connected via the bulk insulating bands and give the anomaly. We will show below that our Rashba-bilayer geometry in the AF phase mimics the 2+1 D surface states of a 3+1 D AF insulator, but with important differences. The realized AF topological insulators break the translational symmetry and protected by a poseudo-TR symmetry which combined broken TR and AF wavevector.\cite{MnBiTe} Our present AF and SV phases do not break any translational symmetry and there arises no such pseudo-TR symmetry. The two Rashba monolayers with opposite SOC and opposite AF orders exhibit $\pm 1/2$ Hall conductivity when the inter-layer tunneling $D_{\bf k}=0$. But for finite $D_{\bf k}$, the conductivity is no longer quantized.

We define a different invariant here. With a proper rotation of the spinor, we can obtain a hidden chiral vortex like structure in the underlying Hamiltonian in both AF and SV phases, whose phase stiffness gives an chiral vector field, and hence a magneto-electric coupling as an anomaly indicator.  

\subsubsection{AF order}
The AF perturbation $H_{\rm AF}$ to $H_0$ acts as a FM exchange energy for each layer, but with opposite spin polarizations between the two layers (we remind that by layer we mean the $h_{A/B}$ blocks in Eq.~\eqref{Ham1}, not the block-diagonal $h^{\pm}$ forms in Eq.~\eqref{BlockH1}). Such a magnetic order is recently observed in CrI3 bilayer system with unusual magneto-electric coupling.\cite{CrI3ME} In the bonding, anti-bonding basis, the AF perturbation mixes the two blocks $h^{\pm}_{\bf k}$, and thus a Chern number can no longer be defined. The bulk band structure, shown in Fig.~\ref{fig_AF}(a), is adiabatically connected to $H_0$, in both non-trivial ($0<D_0<1$), and trivial topological phases. Our explicit TR polarization calculation of the sewing matrix yields $\nu=1$ despite the loss of TR symmetry, see the phase diagram in Fig.~\ref{fig_AF}(c).

Few important properties of this system can be obtained starting with the limit $D_{\bf k}\rightarrow 0$. Here, the two layers, with opposite magnetic orders, become free from electron tunneling, and are only related to each other by interaction. The corresponding Hamiltonian splits into two blocks (from Eq.~\eqref{Ham1}):   
 \begin{eqnarray}
H_0 + H_{\rm AF} &=& 
\begin{pmatrix}
	h_{\rm A} + E_{\rm AF}\sigma_z &  {\bf 0}    \\
	{\bf 0}   &  h_{\rm B} - E_{\rm AF}\sigma_z \\
\end{pmatrix},\label{eq:Ham_AF}
\end{eqnarray}
where ${\bf k}$-dependence is suppressed for simplicity. Recall that $h_{\rm A,B}=\pm {\bm \alpha}_{\bf k}\cdot{\bm  \sigma}$, both layers possess opposite helicity. Each block in Eq.~\eqref{eq:Ham_AF} gives a half-integer Hall conductance $\sigma_{xy}^{\rm A}=-\sigma_{xy}^{\rm B}={\rm sgn}(E_{\rm AF})e^2/2h$. Hence the system has a zero net charge conductance $\sigma_{xy}^{\rm tot}=0$, but an integer-valued spin Hall conductance $-$ reflecting the \ZZ - invariance of the bulk system. This is also termed as half-plateau QAH state in which one expects to observe a zero-Hall conductivity plateau ($\sigma_{xy}^{\rm tot}=0$) as a function of applied magnetic field.\cite{TMEZhang} Therefore $h_{\rm A/B}$ in the limit of $|D_{\bf k}|<<|E_{\rm AF}|$ exactly mimics the the top and bottom surface states of a TR invariant 3D topological insulator, which opposite spin-polarization.

As $D_{\bf k}$ is turned on, the Hall conductance is no longer quantized and becomes a dynamical variable. We find a suitable basis where $D_{\bf k}$, and $E_{\rm AF}$ is combined in a complex-inter-species hopping $-$ an analog of the flux dependent hopping in the twisted boundary condition. We consider a complex extension of the bonding and anti-bonding states as $\psi_{b/a,\sigma}=(\psi_{{\rm A}\sigma} \pm i \psi_{{\rm B}\sigma})/\sqrt{2}$. We choose a spinor $\Psi$=($\psi_{a\uparrow}$, $\psi_{b\downarrow}$, $e^{-i\pi/2}\psi_{b\uparrow}$, $e^{i\pi/2}\psi_{a\uparrow}$)$^T$, governed from the original basis by an unitary rotation $U=\mathbb{I}_4 -i\tau_x\otimes\sigma_z$. In this case, we find that $U(H_0+H_{\rm AF})U^{-1}=H_0'$, where $H_0'$ is the same as the non-interacting Hamiltonian [Eq.~\eqref{Ham1}], except the inter-layer hopping element is now complex: $D'_{\bf k}=D_{\bf k}+ i E_{\rm AF}$. The form of the $H_0'$ with two complex hopping terms, namely SOC coupling $\alpha_{\bf k}$ and $D'_{\bf k}$ resembles the Jackiw-Rossi model in 2D\cite{JachieRossi}. In the low-energy limit, the SOC term governs the Dirac-like fermionic excitations, and the complex $D'_{\bf k}$ represents a vortex. The coupling of a fermion with vortex is the building block of anyons in which their exchange statistics gives a phase other than (0 or $\pi$ ) originating from winding around the vortex. 


Here we are interested in the `persistent' current associated with the phase gradient of the vortex. The phase of the vortex is $\theta_{\bf k}={\rm tan}^{-1}(E_{\rm AF}/D_{\bf k})$. We recall that the band inversion is related to the sign-reversal of $D_{\bf k}$, which in term gives a sign reversal of the phase $\theta_{\bf k}$.\cite{footvortex} In fact, the phase changes discontinuously between $\pm \pi$ across the nodal ring of $D_{{\bf k}_0}=0$.  The discontinuous jump of $\theta_{\bf k}$ on the nodal ring of radius $k_0$ gives a contour of singularity in the phase gradient $\nabla\theta_{\bf k}=\pi\delta(k-k_0)\hat{k}$. Hence winding number of the phase in 2D BZ which gives the anomalous (dynamical) Hall conductivity:
\begin{eqnarray}
\sigma_{xy}^{\pm} &=& \pm\frac{e^2}{\pi h}\frac{1}{(2\pi)^2}\int d{\bf k}\cdot \nabla\theta_{\bf k}=\pm\frac{e^2k_0}{2\pi h}
\label{Eq:AHE}
\end{eqnarray}
where $\pm $ denote the two complex bonding/anti-bonding states. This is the axion analog of the dynamical \ZZ~anomaly of the AF state in 2+1 dimension, but its not an axion invariant.

\subsubsection{SV order}

A novel TR breaking perturbation can be achieved without any magnetic moment through a complex sublattice vortex (SV) order. The corresponding perturbation term $H_{\rm SV}$ is given in Eq.~\ref{Eq:Perturbation} within the mean-field approximation. In the total Hamiltonian $H_0+H_{\rm CS}$, the perturbation term is absorbed into a complex inter-layer hopping as $D'_{\bf k} = D_{\bf k}+ i E_{\rm CS}$. Therefore, this Hamiltonian exactly maps to the AF Hamiltonian in the form $H_0'$ discussed in the AF case above. Because of this vortex like complex hopping, we refer to this state as Sublattice Vortex (SV) state, which is analogous to the AF phase, but have few different symmetry properties. 

Note that despite the same form of the Hamiltonian, the symmetry is different in the two cases since the spinor is different between them. Due to complex $D_{\bf k}'$, SV state breaks \P~ and \T, but is \PT-invariant, as in the AF phase. However, unlike the AF state, the SV state breaks \C, and \CP, and is invariant under sublattice symmetry \CT. Therefore, according to {\it ten-fold way} classification, the system belongs to the AIII group, as in SM phase. However, given that the SV phase is adiabatically connected to the AF phase, both states feature the same topological properties as deduced in Eq.~\ref{Eq:AHE}.

\section{Discussion and outlook}\label{Sec:Discussion}

Our results of the persistence of symmetry-constrained topological phases even after the spontaneous loss of the symmetry are counter-intuitive at first sight. Spontaneous symmetry breaking occurs due to many-body effects in an interacting system when the system's action moves to a new classical ground state. In such a description, the symmetry is actually lost in the wavefunction, not in the original Hamiltonian. Due to the ground state's classical nature, there arise quantum fluctuations that tend to be dominant near the critical point, but is typically superseded by the quasiparticle energy gap in the ordered phase.

Focusing on the latter regime, we can now construct a mean-field Hamiltonian to parameterize the classical ground state consistent with the broken symmetry (s). In the present case, in all the mean-field Hamiltonians, the spinor dimension remains the same, since no other spatial transnational or rotational symmetry is broken. This puts us in an advantageous position to adiabatically bridge the mean-field Hamiltonian to the non-interacting Hamiltonian. In all the ordered phases, simply a gap term arises which couples different spin/sublattices of the original species. In this respect, all the conclusions drawn in this work are also applicable to the explicit symmetry breakings by external electro-magnetic fields or perturbations.

In this mean-field prescription of the symmetry breaking phases, although none of the space or point group symmetries are broken, but the internal symmetry of the Hilbert space plays an important role. We have emphasized throughout the manuscript that as the Hilbert space dimension increases above a critical number (usually $N=$ 2 or 3), the winding number classification changes from \Z~ to \ZZ class. In our non-interacting case as well as in the FM and SM cases, the four-component ($N=4)$ Hilbert space has a  $N=2$ irreducible representation. In 2+1 d, the SU(2) internal symmetry dictates a \Z~class winding number, and hence the Chern number is defined. Therefore, although the system exhibits a \ZZ~ topological class, but it inherently consists of two \Z~classes with opposite winding numbers. For the case of SD, AF and SV phases, there is no $N=2$ component irreducible representation available and our obtained topological invariant for $N=4$ should be consistent with the {\it ten-fold} classification case. In the SD phase, the consistency is bridged. However, for the AF and SV phases, we observe that even if the TR symmetry is lost, but an equivalent $(\mcpt)^2=-1$ or $(\mccp)^2=-1$ or $(\mccpt)^2=-1$ is invariant. More interestingly, the TR polarization still gives a quantised value in the these two phases even after the TR symmetry is broken.

Let us revisit some of the key findings of this work. (i) Despite the loss of protecting symmetries, the topological invariants continue to exhibit the same values, with the same edge state properties. Notably, with the TR symmetry loss, the edge states remain gapless at the TR-invariant ${\bf k}$-point. (ii) According to the symmetry-wise classification scheme in the \textit{ten-fold~way} table for $N>2$ dimensional Hilbert space in 2+1 d, the  change in symmetry is supposed to shuffle the topological groups between \Z, \ZZ, and 0. On the other hand, as long as the original Dirac mass is not closed, we find both the topological properties and its topological groups remain the same for lower dimensional Hilbert space. (iii)  In fact, the TR symmetric sewing matrix, which uniquely defines the TR polarization index $\nu$, continues to be the same even after the loss of the TR symmetry. Only when the band gap closes, the sewing matrix itself becomes zero, and the formula becomes inapt. How do we understand these puzzles? 

One important hint to understand the above phenomena may be the 't Hooft anomaly,\cite{tHooft} as briefly mentioned in Sec~\ref{Sec:Prelude}. As such topology arises from the obstruction to the global gauge symmetry in the manifold the system lives it. Let us take the example of the Ahronov-Bohm geometry, or the quantum Hall droplet in which the electron experiences a curl-free vector potential on its path, but the path encircles a region which inherits a finite magnetic field in the interior. A singular gauge describes such an irrotational vector potential, and any trivial gauge transformation can not gauge it out. In the momentum space, the analog situation arises in certain Hamiltonian when its eigenstates live in hypersphere $\mathbb{S}^d$, where one also cannot uniquely fix the gauge at all points (this is termed as an obstruction to global gauge fixing). In other others, there arises an intrinsic singular gauge field which is the Berry connection.    

But a violation of the global symmetry is a problem to our quantum theory since it is related to the global charge conservation. So, the way we fix this issue is by finding another relevant symmetry to break. This way the `charge' conservation appear to be broken when we individually look at the two states transformed under this symmetry, but globally the total charge is conserved. For example, the axial/chiral symmetry is a symmetry of the Dirac Hamiltonian in an odd spatial dimension. For the massless Dirac case, one has degenerate copies of the Dirac excitations related by the chiral symmetry. Now if we can split them in space in such a way that electrons can flow only from one chiral state to another, but not in reverse, we will have more electrons in one chiral state than in another. We obtain such a geometry in integer quantum Hall (IQH) and Chern insulators where the two chiral modes are split in different edges but connected via the insulating bulk bands.  This gives the chiral anomaly, and the corresponding IQH or Chern insulators are said to be protected by the Chiral symmetry which is actually anomalous here. 

Now we consider a chiral symmetry breaking term $m\bar{\psi}_+\psi_-+ {\rm h.c.}$ (where $m$ is chiral mass) which allows an additional direct tunneling between the two chiral modes ($\psi_{\pm}$). However, the mass $m$ term does not close the bulk insulating gap, and hence the original chiral anomaly phenomenon persists. Therefore, the chiral symmetry breaking term does not immediately destroy the topology until the bulk gap is closed.  This phenomenon is known in the literature of the chiral anomaly.\cite{ChiralBroken,ChiralBroken2}

Generalization of the above discussion to the case of TR symmetry in which the anomaly arises between the two Kramer's partners is obvious. This is, in fact, what has happened in our examples. We have considered those examples of TR symmetry breaking perturbations that do not destroy the bulk Dirac mass immediately. Thus they do not vitiate the \textit{anomaly inflow} between the two Kramer's partners. Hence both the gapless Kramer's partners at the edge as well as TR polarization in the bulk remains the same until the bulk gap is closed.  

However, when it comes to classifying the symmetry of the eigenstates, the TR symmetry is broken, and one cannot anymore use the mathematical mapping to the coset group by using the TR symmetry. Hence the mathematical mapping of the coset group to the homotopy group is lost. On the other hand, as the electron wraps around the BZ ($\mathbb{T}^2$), it continues to experience both $\mathbb{T}^2_{\pm}$ hemispheres which amount to clockwise and anti-clockwise wrapings in $\mathbb{S}^1$. The TR breaking perturbation has acted on adding an innocuous trivial gauge field to the Berry connection that does nor affect the topology. This phenomenon of symmetry breaking is opposite to the theory of symmetry anomaly. In the symmetry anomaly case, the symmetry is intact in the classical theory, but is broken (anomalous) at the quantum level. In the present case, the symmetry is lost at the classical theory, but until the perturbation crosses some critical value, the quantum anomaly does not sense it. 

This can be rigorously proved by going to the long-wavelength limit of the mean-field Hamiltonian, in which one obtains a low-energy action of Dirac field with a Dirac mass and a TR breaking mass. By integrating out the fermion fields, we can obtain an effective Chern-Simon's term owing to the anomaly in the TR symmetry. 
The anomaly persists as long as the TR breaking mass is lower than the Dirac mass.\cite{ChiralBroken,ChiralBroken2} This theory is similar to what is done for the massive chiral symmetry in the high-energy physics, and it will be investigated in future studies. 

\begin{acknowledgments} 
We thank Ramal Afrose for the help with the calculation of the TR polarization. We are thankful to the S.E.R.C. computational facility at the Indian Institute of Science. TD acknowledge research funding from S.E.R.B. Department of Science and Technology, India under I.R.H.P.A grant (File no: IPA/2020/000034.
\end{acknowledgments}

\appendix

\appendix

\section{Gamma Matrices}\label{AppendixA}

The first five gamma matrices are : $\g_1=\tau_z\otimes\sigma_x$, $\g_2=\tau_z\otimes\sigma_y$, $\g_3=\tau_z\otimes\sigma_z$, $\g_4=\tau_x\otimes\sigma_0$ and $\g_5=\tau_y\otimes\sigma_0$. $\tau_i$ and $\sigma_i$ are the $2\times 2$ Pauli matrices in the sublattice and spin space, respectively. These are the generators of the Clifford algebra and satisfy,
\begin{equation}
\{\g_a,\g_b\}=2\delta_{ab}\quad a,b=1,2,\dots,5.
\end{equation}$\g_0=\tau_0\otimes\sigma_0$  and all the other gamma matrices are defined as 
\begin{equation}
\g_{ab}=\frac{1}{2 i}\left[\g_{a}, \g_{b}\right],\quad a,b=1,2,\dots,5.
\end{equation}The explicit form of all the Gamma-matrices are given in Table~\ref{table:4} for convenience.

\begin{table}[t]
\renewcommand{\arraystretch}{1}
\centering
\begin{tabular}{|c|c|}
\hline
$\g_0$ & $\tau_0\otimes\sigma_0$\\
\hline
$\g_1$ & $\tau_z\otimes\sigma_x$\\
\hline
$\g_2$ & $\tau_z\otimes\sigma_y$\\
\hline
$\g_3$ & $\tau_z\otimes\sigma_z$\\
\hline
$\g_4$ & $\tau_x\otimes\sigma_0$\\
\hline
$\g_5$ & $\tau_y\otimes\sigma_0$\\
\hline
$\g_{12}$ & $\tau_0\otimes\sigma_z$\\
\hline
$\g_{13}$ & $-\tau_0\otimes\sigma_y$\\
\hline
$\g_{14}$ & $\tau_y\otimes\sigma_x$\\
\hline
$\g_{15}$ & $-\tau_x\otimes\sigma_x$\\
\hline
$\g_{23}$ & $\tau_0\otimes\sigma_x$\\
\hline
$\g_{24}$ & $\tau_y\otimes\sigma_y$\\
\hline
$\g_{25}$ & $-\tau_x\otimes\sigma_y$\\
\hline
$\g_{34}$ & $\tau_y\otimes\sigma_z$\\
\hline
$\g_{35}$ & $-\tau_x\otimes\sigma_z$\\
\hline
$\g_{45}$ & $\tau_z\otimes\sigma_0$\\
\hline
\end{tabular}
\caption{$\Gamma$ matrices.}
\label{table:4}
\end{table}

\section{Extension to higher dimensional Hamiltonian}\label{AppendixB}
Given that the topological classification table depends on the Hilbert space dimension, we investigate how our findings are sensitive to a higher dimensional Hamiltonian. We add the following trivial part $H_{\rm tr}$ and a coupling term $H_{i,{\rm tr}}$ to the $4\times 4$ non-trivial Hamiltonian $H_{i}$, where  $i$ stands for various Hamiltonians we considered in the main text.
\begin{eqnarray}
H_{2} &=& 
\begin{pmatrix}
	H_{i} &  H_{i,{\rm tr}} \\
	H^{\dag}_{i,{\rm tr}}   &   H_{{\rm tr}}\\
\end{pmatrix}.\label{eq:ExtendedHam}
\end{eqnarray}
To keep the symmetry of the entire system as tht of the non-trivial part $H_i$, we choose the basis of the trivial Hamiltonian as $(\psi_{C\uparrow},~ \psi_{C\downarrow},~...)^T$, where $C$ is a third single Rashba layer, and so on. A generic choice would be $H_{{\rm tr}}=(h_{{\rm C},{\bf k}}, ...$, where $h_{{\rm C},{\bf k}}$ is the same Rashba Hamiltonian used in Eq.~\eqref{Ham1}. $H_{i,{\rm tr}}$ is also kept to the spin-conserving hopping as in the main text. If we set $H_{i,{\rm tr}}=D_{\bf k}$, it was shown in Ref.~\cite{DasRashba} that above roughly six Rashba-bilayers (with opposite SOC) a 3D \ZZ~ topological insulator phase commences in which the top and bottom Rashba single layers host gapless 2D Dirac cones.

To avoid a 3D generalization, here we simple consider $H_{i,{\rm tr}}=D_{\bf 0}$, i.e., we keep a momentum-independent coupling term for $H_{i,{\rm tr}}$ by setting $D_1=0$. Then it is easy to understand that the topology of the entire Hamiltonian is same as that of $H_{i}$. For an even number of Rashba single-layers, say, $8\times 8$ total Hamiltonian, Rashba-bilayer with opposite SOCs construction holds and we can block and off-block diagonalize the entire Hamiltonian. Therefore, the entire analysis of Chern number, parity inversion, \T-polarization, particle-hole anomaly, wherever applicable in the main text, also works here. For an odd-number of Rashba single-layers, the bottom most Rashba layer (`C' layer) does not have its opposite Rashba SOC counter part. Therefore, the total Hamiltonian may no longer be block or off-block diagonalized, and hence the Chern number description does not work. The system also does not have the parity symmetry (even if $H_i$ has). However, our explicit \T-polarization calculation shows that all results remain the same to that of $H_i$. Hence, we conclude that our findings are insensitive to the Hilbert space dimension as long as an irreducible $SU(2)$ internal symmetry exists.


\begin{thebibliography}{1}
\bibitem{foot1}The anomaly here is a consequence of deriving a classical low-energy theory, which disappears in the full quantum theory. Because, when one writes a low-energy theory at one boundary, there exists only one symmetry partner, and theory would appear to have lost the symmetry. However, in a full lattice theory including both boundaries and bulk states, both symmetry partners are present and the symmetry is intact.
\bibitem{footCPT}\CPT~invariance in high-energy physics is associated with Lorentz invariance, which can be an emergent symmetry here, however, the definition of the chiral symmetry is not necessarily the same, and hence we do not make any such analogy for the \CPT~ invariance case.
\bibitem{foot2}We simple refer here that all values of the unitary matrix $U$ at different momenta are isomorphic to all points in a hypershere $\mathbf{S}^d$.
\bibitem{footexample} For example, imagine a two-component vector bundle defined on the tangent space of a M\"obius band.  It is easy to notice that in a M\"obius band, a vector undergoes a rotation by $\pi$ as in completed one rotation, while on a trivial cylinder surface there is no winding of the vectors.
\bibitem{footLargeGauge} For the Abelian $U$(1)  gauge transformation, since $\Omega_{\bf k}$ is unitary, it can be expressed by a phase as $\Omega_{\bf k}=e^{i\chi_{\bf k}}$, where $\chi_{\bf k}$ is real number, which is what Kane-Mele introduced in  Ref.~\onlinecite{KaneMele} For a non-Abelian case, $\Omega_{\bf k}$ belong to a $SU(N)$ unitary group where $N$ is the degeneracy of each $\psi_{\pm}$ state.
\bibitem{FootGaugefixing}In numerical simulation, one needs to make an additional gauge fixing to remove the other innocuous gauge fields (also known as small gauge transformation) which are not related to the topology [See Ref.~\onlinecite{Vanderbilt}].
\bibitem{foot_PH} We note that throughout this paper we refer particle-hole symmetry, if present, to the energy spectrum only, which can arise due to various symmetries anticommuting with the Hamiltonian. Unlike some other papers, the presence of particle-hole symmetry does not necessarily imply the presence of charge conjugation symmetry.
\end{thebibliography}
\end{document}